\documentclass[prr,twocolumn,showpacs,amsmath,amssymb,superscriptaddress,floatfix,reprint, nofootinbib]{revtex4-1}
\usepackage{amsmath}
\usepackage{amssymb}
\usepackage{amsthm}
\usepackage{amsfonts}
\usepackage{dsfont}
\usepackage{listings}
\usepackage{enumerate}
\usepackage{latexsym}
\usepackage{ulem}
\usepackage{hyperref}

\usepackage{bbold}

\usepackage{psfrag}
\usepackage{xcolor}
\usepackage{float}

\usepackage{comment}

\usepackage{graphicx}
\usepackage{subfigure}
\newcommand{\dx}{\Delta x}
\newcommand{\dt}{\Delta t}

\newcommand{\beq}{\begin{equation}}
\newcommand{\eneq}{\end{equation}}
\newcommand{\eeq}{\end{equation}}

\newcommand{\prodal}[2]{\underset{#1}{\overset{#2}{\prod}}}
\newcommand{\bigotimesal}[2]{\underset{#1}{\overset{#2}{\bigotimes}}}
\newcommand{\bigoplusal}[2]{\underset{#1}{\overset{#2}{\bigoplus}}}
\newcommand{\sumal}[2]{\underset{#1}{\overset{#2}{\sum}}}
\newcommand{\intal}[2]{\underset{#1}{\overset{#2}{\int}}}

\newcommand{\bra}[1]{\left\langle#1\right|}
\newcommand{\ket}[1]{\left|#1\right\rangle}

\newcommand{\half}{\frac{1}{2}}

\newcommand{\mE}{\mathcal{E}}
\newcommand{\mk}{\mathcal{K}}

\newcommand{\mC}{\mathcal{C}}
\newcommand{\mZ}{\mathcal{Z}}
\newcommand{\mD}{\mathcal{D}}

\newcommand{\ms}{\mathcal{S}}
\newcommand{\mB}{\mathcal{B}}
\newcommand{\mG}{\mathcal{G}}
\newcommand{\mQ}{\mathcal{Q}}

\newcommand{\vx}{\vec{x}}
\newcommand{\vy}{\vec{y}}
\newcommand{\vk}{\vec{k}}

\newcommand{\hM}{\widehat{M}}
\newcommand{\hW}{\widehat{W}}
\newcommand{\hU}{\widehat{U}}
\newcommand{\hS}{\widehat{S}}
\newcommand{\hQ}{\widehat{Q}}
\newcommand{\hm}{\widehat{m}}

\newcommand{\phim}{\phi^{(m)}}
\newcommand{\Em}[1]{E^{{#1}_0 \cdots {#1}_m}}
\newcommand{\zm}[1]{\zeta^{{#1}_0 \cdots {#1}_m}}

\newcommand{\mbP}{\mathbb{P}}
  
\input{epsf}

\newcommand{\Tr}{\mathrm{Tr}}
\newcommand{\tth}{t_\mathrm{Th}}
\newcommand{\hrk}{H_\mathrm{RK}}
\newcommand{\hsmf}{H_\mathrm{SMF}}
\newcommand{\grk}{\ket{G_{\mathrm{RK}}}}

\newcommand{\nn}{\nonumber}

\newcommand{\twopartdef}[4]
{
	\left\{
		\begin{array}{ll}
			#1 & \mbox{if } #2 \\[5mm]
			#3 & \mbox{if } #4
		\end{array}
	\right.
}

\begin{document}

\tolerance 10000
\title{Spectral statistics in constrained many-body quantum chaotic systems}
\author{Sanjay Moudgalya}
\affiliation{Department of Physics, Princeton University, Princeton, NJ 08544, USA}
\affiliation{Department of Physics and Institute for Quantum Information and Matter,
California Institute of Technology, Pasadena, California 91125, USA}
\affiliation{Walter Burke Institute for Theoretical Physics, California Institute of Technology, Pasadena, California 91125, USA}
\author{Abhinav Prem}
\affiliation{Princeton Center for Theoretical Science, Princeton University, Princeton, NJ 08544, USA}
\author{David A. Huse}
\affiliation{Department of Physics, Princeton University, Princeton, NJ 08544, USA}
\author{Amos Chan}
\affiliation{Princeton Center for Theoretical Science, Princeton University, Princeton, NJ 08544, USA}
\begin{abstract}
We study the spectral statistics of spatially-extended many-body quantum systems with on-site Abelian symmetries or local constraints, focusing primarily on those with conserved dipole and higher moments. In the limit of large local Hilbert space dimension, we find that the spectral form factor $K(t)$ of Floquet random circuits can be mapped exactly to a classical Markov circuit, and, at late times, is related to the partition function of a frustration-free Rokhsar-Kivelson (RK) type Hamiltonian. Through this mapping, 
we show that the inverse of the spectral gap of the RK-Hamiltonian lower bounds the Thouless time $\tth$ of the underlying circuit. For systems with conserved higher moments, we derive a field theory for the corresponding RK-Hamiltonian by proposing a generalized height field representation for the Hilbert space of the effective spin chain. Using the field theory formulation, we obtain the dispersion of the low-lying excitations of the RK-Hamiltonian in the continuum limit, which allows us to extract $\tth$. In particular, we analytically argue that in a system of length $L$ that conserves the $m^{th}$ multipole moment, $\tth$ scales subdiffusively as $L^{2(m+1)}$. We also show that our formalism directly generalizes to higher dimensional circuits, and that in systems that conserve any component of the $m^{th}$ multipole moment, $\tth$ has the same scaling with the linear size of the system. Our work therefore provides a general approach for studying spectral statistics in constrained many-body chaotic systems.
\end{abstract}
\date{\today}
\maketitle
%


\tableofcontents

\section{Introduction}
\label{sec:intro}
Recent years have seen a surge of interest in understanding the foundations of quantum statistical mechanics.
A convergence of experimental progress in the engineering and manipulation of ultracold atomic gases, which provide excellent examples of isolated quantum systems~\cite{blochrmp}, and profound theoretical insight has brought to the forefront of contemporary research the nature of closed quantum many-body systems evolving under unitary dynamics.
Research in this direction has unearthed a plethora of novel non-equilibrium phenomena, such as many-body localization~\cite{gornyi2005, baa2006, pal2010many, nandkishore2015many}, quantum many-body scarring~\cite{shiraishi2017systematic, moudgalya2018a, turner2018, moudgalya2018b, turner2018quantum, ho2018periodic, khemani2019int}, and Hilbert space fragmentation~\cite{sala2020ergodicity, khemani2020, rakovszky2020statistical, moudgalya2019krylov}, which provide examples of non-integrable interacting systems which fail to obey the Eigenstate Thermalization Hypothesis (ETH)~\cite{deutsch1991quantum, srednicki1994chaos, rigol2008thermalization, polkovnikov2011colloquium, d2016quantum}. Concurrently, ideas from the dynamics of black holes have led to new perspectives on characterizing quantum chaos and diagnostics thereof~\cite{Kudler-Flam2020}, including the decay of Out-of-Time-Order Correlators~\cite{kitaev2015simple,maldacena2016bound,maldacena2016remarks, cotler2017black, cotler2017chaos} and operator spreading~\cite{nahum2017, keyserlingk2018, zanardi2012,hamma2012, kos2018,gharibyan2018, moudgalya2019operator, jonay2018coarsegrained, khemani2018, rakovszky2018}.
These diagnostics complement familiar signatures of quantum chaos derived from the eigenvalue spectrum of Hamiltonian or Floquet systems, such as level repulsion~\cite{bohigas1984characterization, montambaux1993quantum, poilblanc1993poisson} and the Spectral Form Factor (SFF)~\cite{Haake1991}. These ideas rely on the widely-held belief that dynamics of generic many-body quantum chaotic systems beyond a timescale $\tth$, dubbed the ``Thouless time," follow predictions from Random Matrix Theory (RMT)~\cite{thouless1977maximum} i.e., their late-time behavior resembles that of a random matrix chosen from an ensemble consistent with the system's symmetries~\cite{dyson1962brownian, Haake1991, cotler2017chaos}.
%


\begin{figure}[b]
\centering
\includegraphics[scale=2]{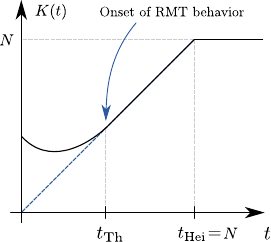}\caption{(Color online)  Illustration of the SFF versus time on a log-log scale. 
The dashed blue line is the CUE RMT behavior. The black solid line is the generic behavior for many-body quantum chaotic systems, characterized by the Thouless time $t_\mathrm{Th}$ and the Heisenberg time $t_\mathrm{Hei}$, the latter of which scales exponentially with system size $L$. The early time behaviour of $K(t)$ depends on the details of the system, particularly on whether the system is defined by a Floquet operator or a Hamiltonian.
}
\label{fig:t_th}
\end{figure}

\begin{figure*}[ht]
    \centering
    \includegraphics[width=0.9 \textwidth]{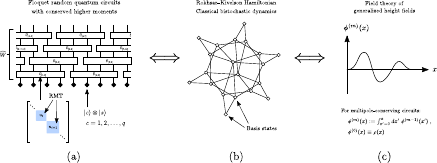}
    \caption{Summary of results: a) We study a class of constrained Floquet random quantum circuits whose SFF in the large-$q$ limit can be related to the partition function of an RK-Hamiltonian. b) Graph representation of the RK-Hamiltonian, with nodes representing an appropriate set of basis states and links representing possible moves between them.
    c) Generalized height field representation for each spin configuration, with symmetries enforced via boundary constraints. The RK ground state is then written as an equal-weight superposition of height fields (equivalently, as random walks in the space direction). 
    }
    \label{fig:summary}
\end{figure*}


%
Despite the significant difficulty in analytically studying dynamics in generic many-body quantum systems, substantial progress has been made in delineating the dynamics of chaos in random quantum circuits via the two-point spectral form factor $K(t)$, defined in terms of the spectral properties of the evolution operator $\hW$ as
\beq
\label{eq:K_def}
K(t) := \left \langle \sum_{m,n=1}^N e^{i (\theta_m - \theta_n) t }\right\rangle = 
\left\langle \left| \Tr[\hW(t)] \right|^2  \right\rangle 
\; ,
\eeq
where $\{\theta_m \}$ is the set of eigenphases of $\hW$, $\hW(t) \equiv \hW^t$ denotes the $t^{th}$ power of $\hW$, $N$ is the Hilbert space dimension, and $\langle \cdot \rangle$ denotes the average over an ensemble of statistically similar systems. The SFF is the Fourier transform of the two-level correlation function, with time $t$ as the variable conjugate to $\omega \sim \theta_m -\theta_n$, the (quasi)-energy difference. 
For $\hW$ with Poisson-distributed eigen-levels, $K(t)= N$ for all $t$, while for $\hW$ chosen as random matrices from the Circular Unitary Ensemble (CUE), $K(t) = |t|$ (the ``ramp'') for times well below the Heisenberg time $t_{\mathrm{Heis}} = N$, after which it plateaus to $K(t) = N$~\cite{Haake1991}.
The SFF serves as a barometer for quantum chaos, as the appearance of RMT predictions in the SFF provide a crisp time-scale for characterizing the many-body chaos in a finite system, which we take as defining the Thouless time $\tth$\footnote{In systems with diffusive or subdiffusive dynamics, this is expected to be a good estimate of the actual Thouless time, since it is always slow compared to the operator spreading time.}; specifically, we define $\tth$ as the time scale at which the SFF approaches the CUE RMT behavior (see Fig.~\ref{fig:t_th}).
In contrast to nearest-neighbour level spacing distributions, the SFF encodes spectral correlations at \textit{all} time (equivalently, energy) scales and is also relatively simple to analyse, as it involves only two sums over the (quasi)-energy eigenvalues.

One approach for analytically computing the SFF exploits a self-duality present in certain models~\cite{bertini_sff, flack2020}. However, the applicability of this approach is limited only to self-dual circuits and does not extend to generic interacting circuits, possibly with conserved quantities.
A second, more generic approach~\cite{cdc1, cdc2, friedman2019, Chan_2019, garratt2020} studies Floquet random quantum circuits (FRQCs) in the limit of large local Hilbert space dimension, which are amenable to exact analytic calculations of the SFF and hence enable one to study the implications of conserved quantities on dynamics.
For circuits with a globally conserved U(1) charge, $\tth$ was shown to scale diffusively as $\sim L^2$, validating the idea that RMT behaviour is established only once all conserved quantities have diffused through the system.
In contrast, for certain systems without any conserved charge, $\tth \sim \log L$~\cite{cdc2}.
In these latter systems, the SFF is not sensitive to the slower ``ballistic'' dynamics due to locality and causality which means the operator spreading differs from RMT dynamics up to longer times of order $L$.
Besides systems with conserved charges, there has been growing interest in the dynamics of \textit{constrained} non-integrable quantum systems with more general symmetries, driven partly by the discovery of their anomalous dynamics~\cite{chamon2005,kim2016,siva2017,prem2017, chandran2016eigenstate,bernien2017probing, turner2018, turner2018quantum, feldmeier2019emergent,iaconis2019anomalous}, which resembles that of classical Kinetically-Constrained-Models~\cite{ritort2003glassy}.
Of particular interest are systems which conserve both charge and dipole moment (or center-of-mass)~\cite{pai2018localization, pai2019dynamical, sala2020ergodicity, moudgalya2019quantum, moudgalya2019krylov, rakovszky2020statistical,  morningstar2020,feldmeier2020}, symmetries which naturally appear in systems subjected to strong electric fields~\cite{bakr2019, moudgalya2019krylov, khemani2020} and in fracton models~\cite{fractonreview,fractonreview2}.
Dipole moment conserving systems exhibit various novel dynamical phenomena including operator localization~\cite{pai2018localization} and Hilbert space fragmentation~\cite{pai2019dynamical, sala2020ergodicity, khemani2020}, coexistence of integrable and non-integrable subspaces leading to a restricted form of ETH~\cite{moudgalya2019krylov, moudgalya2019quantum}, presence of topological edge modes in highly excited states~\cite{rakovszky2020statistical}, and subdiffusive transport~\cite{gromovhydro,feldmeier2020, zhang2020subdiffusion}, which has been experimentally observed~\cite{bakr2019}.
In this paper, we develop a general approach for studying features of the spectral statistics in constrained quantum chaotic many-body systems, focusing on those with conserved dipole and higher moments.
Using the SFF $K(t)$ as a diagnostic for many-body chaos, we extract the scaling of this $\tth$ with system size $L$ for one-dimensional (1D) FRQCs with a local Hilbert space comprising $q$ `color' degrees of freedom (DOFs) coupled with auxiliary spins through which the constraints are imposed. 
In the large-$q$ ($q \rightarrow \infty$) limit, we express $K(t)$ in terms of a classical Markov circuit which inherits the constraints of the underlying FRQC, as illustrated in Fig.~\ref{fig:summary}(b).
Utilizing an established correspondence between classical Markov processes and Rokhsar-Kivelson (RK) type Hamiltonians, we can equivalently relate $K(t)$ at late times to the partition function of a positive-definite, frustration-free RK-Hamiltonian\footnote{As discussed later in Sec.~\ref{sec:mapping}, we take ``RK-Hamiltonian" to mean a quantum Hamiltonian that is proportional to the transition matrix of a Markov process that satisfies detailed balance. This taxonomy stems from the quantum dimer context~\cite{rkham}, where the terms ``RK-Hamiltonian" and ``quantum dimer model at the RK-point" are often used interchangeably.} acting solely on the spin DOFs. 
Consequently, a lower bound on $\tth$ can be extracted from the spectral gap of the RK-Hamiltonian; this mapping hence allows us to borrow techniques from equilibrium physics to establish dynamical properties of the underlying circuit~\cite{henley1997relaxation, castelnovo2005, somma2007quantum}.

For circuits with conserved higher moments in one dimension, we find a continuum representation for the ground state (GS) of the RK-Hamiltonian in terms of generalized ``height" fields (see Fig.~\ref{fig:summary}(c)), from which we identify a continuum parent Hamiltonian for the corresponding GS and establish a lower bound on $\tth$.
We find a sub-diffusive scaling of $\tth \sim L^{2(m+1)}$ for circuits of length $L$ with a conserved $m^{th}$ moment i.e., the timescale at which random matrix behavior ensues in such systems is parametrically longer than in systems with only a conserved U(1) charge, which spreads diffusively ($m=0$: $\tth \sim L^2$). 
We also find similar results in higher dimensions by constructing continuum representations of the ground state and the RK-Hamiltonian in terms of generalized tensor fields, which predicts $\tth \sim L^{2(m+1)}$ for a system with linear-size $L$ with conserved $m^{th}$ moments in all directions.
Note that while the field theories are specific to higher-moment conserving systems, the mapping from $K(t)$ to an emergent RK-Hamiltonian in the large-$q$ limit holds generally.
This paper is organized as follows. We start by defining the class of FRQCs under consideration in Sec.~\ref{sec:FRQC}. In Sec.~\ref{sec:mapping}, we study these circuits in the limit of large local Hilbert space and establish a mapping between the SFF of the FRQC and the dynamics of a classical Markov chain, which we further show is equivalent at late times to the partition function of an emergent RK-Hamiltonian. Through these mappings, we find that the Thouless time $\tth$ of the underlying circuit is lower bounded by the spectral gap of this emergent Hamiltonian. Focusing on systems with conserved higher moments in Sec.~\ref{sec:multipole}, we verify that $\tth \sim L^2$ in charge conserving systems and provide numerical evidence for subdiffusive scaling of $\tth$ in systems that additionally conserve the global dipole moment. In Sec.~\ref{sec:continuum}, we take the continuum limit of the emergent RK-Hamiltonian for systems which conserve all moments up to the $m^{th}$ highest moment. We extract a bound on $\tth$ from the dispersion relation of the continuum Hamiltonian and show that $\tth \sim L^{2(m+1)}$. In Sec.~\ref{sec:higherdims}, we generalise our results to multipole conserving systems in higher dimensions. We conclude in Sec.~\ref{sec:cncls} with a discussion of open questions and future directions.


\section{Constrained Floquet Random Quantum Circuits}
\label{sec:FRQC}
Our primary object of interest in this paper is the Thouless time $\tth$ of constrained many-body quantum chaotic systems, where we define $\tth$ as the time-scale after which the behaviour of the SFF $K(t)$ closely approaches RMT predictions. To probe $K(t)$, we consider one-dimensional $L$-site spatially-random FRQCs with local Hilbert space at each site of the chain given by $\mathcal{H}_{\text{loc}} = \mathbb{C}^q \otimes \mathbb{C}^{2s+1}$, where $\mathbb{C}^q$ and $\mathbb{C}^{2s+1}$ are the local Hilbert spaces of the \textit{color} and \textit{spin} DOFs respectively. The color DOFs facilitate Haar averaging and allow us to retain analytical control in the $q\to\infty$ limit~\cite{cdc1,cdc2,khemani2018}; the spins, on the other hand, allow us to encode on-site Abelian symmetries, such as U(1) charge conservation (previously considered in Refs.~\cite{khemani2018,friedman2019}), or impose local constraints on the dynamics.
 
More precisely, we consider Floquet circuits defined by a time-evolution operator $\hW$ over a single period composed of unitary gates acting on a finite number $\ell \geq \ell_{\mathrm{min}}$ of contiguous sites, where $\ell_{\mathrm{min}}$ is the ``minimal" gate size for non-trivial local dynamics under the symmetry or constraints of interest. 
Without loss of generality, local dynamics on all sets of $\ell$ contiguous sites within a single time-period can be ensured by choosing $\hW$ to be composed of $\ell$ layers of operators $\{\hW_a\}$, where $\hW_a$ is composed of $r = \lfloor L/\ell \rfloor$ spatially random local unitary gates $\{\hU_{[j,j + \ell - 1]}\}$, and has the form:\footnote{In certain cases, considering Floquet operators with $m < \ell$ layers per period is sufficient to ensure that non-trivial dynamics occurs in all sets of $\ell$ contiguous sites. However, this choice of $m < \ell$ layers leads to identical late-time dynamical features as that of operators with $\ell$ layers.}
\beq
\label{eq:wdef}
\hW = \prod_{a=1}^\ell \hW_a  \, , \quad \hW_a = \bigotimes_{n=1}^r \hU_{[a + (n-1) \ell, a + n \ell - 1]} \, ,
\eeq
where $a$ is the layer index.
As shown in Fig.~\ref{fig:summary}(a), $\hU_{[a + (n-1) \ell, a + n \ell - 1]}$ labels the $n^{th}$ local gate in the $a^{th}$ layer and acts non-trivially only on sites $j \in \{a + n \ell - \ell,\dots, a+ n \ell - 1\}$, where $1\leq n \leq r$ and the site index $j$ is defined mod $L$ for periodic boundary conditions (PBC).
Each of the local gates has the following block-diagonal structure: 
\beq
\label{eq:udef}
\hU_{[j, j + \ell - 1]} = \bigoplus_{\alpha=1}^{\mathcal{D}} u(j,\alpha) \, ,
\eeq
where $\alpha$ denotes each set (block) of $\ell$-site spin-configurations within this gate which are connected through local moves permissible under symmetries or constraints, and $\mathcal{D}$ denotes the total number of such blocks. Block $\alpha$ contains $d_\alpha$ spin configurations within this gate, with $\sum_{\alpha = 1}^{\mathcal{D}} d_\alpha = (2s+1)^\ell$.  Note that we do not impose any constraints on the color DOFs, only on the spins.
Each $u(j,\alpha)$ is thus a $d_\alpha q^\ell \times d_\alpha q^\ell$ unitary drawn independently from the Haar ensemble acting on the states in block $\alpha$, while it gives zero when it acts on all other states.
In particular, for systems without any symmetries or dynamical constraints, the local gates $\hU_{[j, j + \ell - 1]}$ are $(2s+1)^\ell q^\ell \times (2s+1)^\ell q^\ell$ independent Haar random unitaries.
The block diagonal structure of the local gates encodes the symmetries or constraints of interest. Specifically for systems which have global symmetry sectors labelled by a set of quantum numbers $\ms = \{s_1, s_2, \dots \}$, each local $\ell$-site gate $\hU$ is block diagonal, with each block containing all spin states with the same $\ms$. As a technical aside, we note that since we are keeping the gate size $\ell$ fixed while treating all transitions involving those $\ell$ sites on equal footing, this also includes all allowed processes involving spin transitions on any subset of those $\ell$ sites.

As an example, let us consider an FRQC with $s=1/2$ DOFs which preserves the total charge $\hQ_0 = \sum_{x=1}^L \hS^z_x$ of the spins, where $\hS^z_x$ is the Pauli-$Z$ matrix acting on site $x$~\cite{friedman2019}.  To allow non-trivial dynamics, we choose $\ell$ to be equal to $\ell_{\text{min}} = 2$, such that each local gate $\hU$ is a $4q^2 \times 4q^2$ block-diagonal matrix. Each local gate is composed of $\mathcal{D} = 3$ blocks: two $q^2 \times q^2$ blocks act on the tensor product subspaces associated with the spin configurations $\ket{\uparrow \uparrow}$ and $\ket{\downarrow \downarrow}$, and a single $2 q^2 \times 2 q^2$ block acts on the subspace associated with the spin configurations $\ket{ \uparrow \downarrow} , \ket{ \downarrow \uparrow}$. Each of these three blocks locally preserves the U(1) charge over 2-sites and is an independently-drawn Haar random unitary. 

In this paper, we mainly focus on circuits which conserve not only the total charge, but also all higher moments up to the $m^{th}$ moment $\hQ_m = \sum_{x=1}^L x^m \hS^z_x \equiv \sum_{x=1}^L \hQ_m(x)$. Such circuits neatly fall into the larger class of FRQCs defined earlier via Eqs.~\eqref{eq:wdef} and~\eqref{eq:udef}. For instance, for systems with both charge and dipole moment conservation, we can consider $s=1$ DOFs and $\ell = \ell_{\text{min}} = 3$ site gates, where each local gate is a $27q^3 \times 27q^3$ block diagonal matrix, with each block corresponding to those spin configurations which are connected under local (3-site) dipole moment preserving dynamics (see Ref.~\cite{pai2018localization} for details). It is straightforward to generalize the above circuits to higher spins $s$, larger gate sizes $\ell$, and higher moment conservation laws. While our primary focus in this paper will be systems with higher conserved moments, the FRQCs defined in Eqs.~\eqref{eq:wdef} and~\eqref{eq:udef} define a much broader class, including those with arbitrary on-site Abelian symmetries as well as circuits which obey dynamical constraints, such as those present in the PXP model~\cite{bernien2017probing}.

We characterise the spectral features of the above class of FRQCs using the SFF defined in Eq.~\eqref{eq:K_def}. For a circuit $\hW$ invariant under a set of global symmetries, corresponding to a set of operators $\{\hS_1, \hS_2, \dots\}$, we have $[\hW, \hS_i] = 0 \; \forall \, i$. Therefore,
\beq
\hW = \bigoplus_\ms \hW^{(\ms)} 
\label{eq:Wsym}
\eeq
is block-diagonal and quasienergy levels of $\hW$ from blocks $\hW^{(\ms)}$, corresponding to distinct quantum number sectors $\ms = \{s_1, s_2, \dots \}$, do not repel~\cite{d2016quantum}.
In addition, certain systems, such as those with higher moment symmetries, further exhibit the phenomenon of Hilbert space fragmentation, wherein the dynamics does not connect all products states in the $\hS_z$ (equivalently, charge) basis even within the same symmetry sector~\cite{sala2020ergodicity, khemani2020, pai2019dynamical, moudgalya2019quantum}.
As a consequence, within each symmetry sector $\ms$ there may exist up to exponentially many disjoint ``Krylov subspaces," labelled by $\mathcal{K}_i^{(\ms)}$ i.e.,
\beq
\hW^{(\ms)} = \bigoplus_{i=1}^{D^{(\ms)}} \hW^{(\mathcal{K}_i^{(\ms)})},
\label{eq:Wkryl}
\eeq
where $D^{(\ms)}$ denotes the number of disjoint Krylov subspaces generated from product states with the same quantum numbers $\ms$. This fragmented structure of $\hW$ is schematically depicted in Fig.~\ref{fig:frag}.  
Hence, given the possibility of Hilbert space fragmentation in quantum many-body systems, we define the SFF restricted to a given Krylov subspace $\mk$:
\beq
\label{eq:K_krylov}
K\left(t; \mk\right) \equiv \left\langle \left| \Tr_{\mk}[\hW^t] \right|^2\right\rangle 
\; ,
\eeq
where the subscript $\mk$ denotes the restriction of $\hW$ to a Krylov subspace and $\left\langle \cdot \right\rangle$ denotes averaging over the Haar random unitaries in the FRQC.  Note that this definition encompasses systems with global symmetries but no fragmentation, since in that case, each Krylov subspace $\mathcal{K}^{(\ms)}$ fully spans its global symmetry sector $\ms$ and all $D^{(\ms)}=1$.
%
%
\section{Mapping to Classical Markov Chain and Emergent RK-Hamiltonian}
\label{sec:mapping}
\begin{figure}[t!]
\centering
\includegraphics[scale = 1]{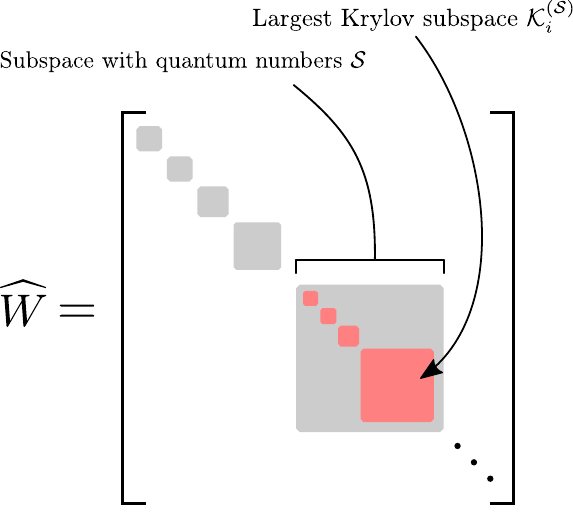}    %
\caption{(Color online) Example of a Floquet operator exhibiting Hilbert space fragmentation in the $Z$-basis. Symmetry sectors are denoted by $\ms$, and Krylov subspaces $\mk$ within symmetry sectors are denoted by $\mathcal{K}_i^{(\ms)}$.}\label{fig:frag}
\end{figure}

Computing the SFF Eq.~\eqref{eq:K_krylov} for many-body quantum systems is analytically difficult in general. Refs.~\cite{cdc1,cdc2,friedman2019} developed a diagrammatic approach for evaluating the ensemble averaging in~\eqref{eq:K_krylov}, by effectively ``integrating out" the color DOFs for random quantum circuits with charge conservation.
In Appendix~\ref{app:RKmapping}, we generalize this technique to the general class of constrained FRQCs discussed in the previous section and find that, to leading order in the large-$q$ limit,
\begin{equation}
\label{eq:K_RK}
    K_\infty(t;\mk) \equiv \lim_{q \to \infty} K(t; \mk) = |t|\ \textrm{Tr}_{\mk} \left[\hM^t\right] \, ,
\end{equation}
where the factor of $| t |$ stems from $| t |$ leading order diagrams as $q\to \infty$ \cite{cdc1}. The Markov matrix $\hM$ is a \textit{classical} bi-stochastic\footnote{An $N\times N$ non-negative matrix $M$ is called bi-stochastic if $\sum_{i=1}^N M_{i,j} = 1\,\forall\,j$ and $\sum_{j=1}^N M_{i,j} = 1\,\forall\,i$.} circuit acting only on effective spin-$s$ DOFs and is composed of local $\ell$-site gates.
Furthermore, $\hM$ inherits the circuit geometry, symmetries, and Krylov subspaces of $\hW$ and can be expressed as
\beq
\hM = \prodal{a = 1}{\ell}{\hM_a} \, , \quad \hM_a = \bigotimesal{n = 1}{r}{\ \hm_{[a + (n-1)\ell, a + n \ell - 1]}} \, 
\eeq
where, in analogy with Eq.~\eqref{eq:wdef}, $a$ is the layer index and $\hm_{[a + (n-1)\ell, a + n \ell - 1]}$ labels the $n^{th}$ local gate in the $a^{th}$ layer, acting on sites $j \in \{a + n \ell - \ell, \dots, a + n\ell -1 \}$.
As before, $1\leq n \leq r$ and $j$ is defined mod $L$ for PBC. However, unlike the underlying gates $\hU$, the $\ell$-site gates $\{\hm_{[j, j + \ell - 1]}\}$ are \textit{non-random}:
\beq\label{eq:m_gates}
\hm_{[j, j + \ell - 1]} = \bigoplus_{\alpha=1}^{\mathcal{D}} m(d_\alpha), \quad  
m(d_\alpha) =\frac{1}{d_\alpha}
\begin{bmatrix}
1 & 1 & \dots
\\
1 & 1 & \dots
\\
\vdots & \vdots & \ddots 
\end{bmatrix}_{d_\alpha \times d_\alpha}
\; ,
\eeq
where the $d_\alpha$'s are the sizes of the blocks of $\ell$-site spin configurations that are dynamically connected and $\mathcal{D}$ is the number of dynamically connected blocks in $\hm$. 
The fact that Eq.~\eqref{eq:m_gates} retains the block-diagonal form with equal matrix elements (within each block) is consistent with the fact that the dynamical constraints are imposed via the spin DOFs and that the local Haar random gates are invariant upon a change of basis. 
Since $\hM$ inherits the symmetries and Krylov subspaces (Eqs.~\eqref{eq:Wsym} and~\eqref{eq:Wkryl}) of $\hW$, and  $\hM$ also has a block diagonal structure (see Fig.~\ref{fig:frag}), where each block is itself an irreducible bi-stochastic matrix; hence, each block has a unique largest magnitude eigenvalue $1$. 
Restricting our attention to the subspace $\mk$ of interest, let us denote the eigenvalues of the corresponding block by $\{\Lambda_j^{(\mk)} \}$, with $\Lambda_1^{(\mk)} = 1$ and ordered such that $|\Lambda_j^{(\mk)}| \geq |\Lambda_{j+1}^{(\mk)}| \, \forall \, j$.\footnote{Note that the eigenvalues of $\hM$ can be complex since is not a symmetric matrix due to the brick-wall structure of the circuit} 
We can then write
\beq
\label{eq:Kexpand}
K_\infty(t; \mk) = |t|\ \textrm{Tr}_{\mk} \left[\hM^t\right]
     = |t|\left(1 + \sum_{j>1} \left( \Lambda_j^{(\mk)} \right)^t \right) \, .
\eeq
At sufficiently long times\footnote{Technically, we require that $t_H \gg t \gg 1$, but the so-called Heisenberg time $t_H$ proportional to the inverse level spacing is infinite in the large-$q$ limit.} $t \gg 1$, we can then expand Eq.~\eqref{eq:Kexpand} as
\beq
\label{eq:Ktexpansion}
K_\infty(t; \mk) = |t|\left(1 + d^{(\mk)} \exp\left(- t \, \Delta^{(\mk)}_{\hM}\right) + \cdots \right) \, ,
\eeq
where $|\Lambda_2^{(\mk)}| = \exp \left( -\Delta^{(\mk)}_{\hM} \right)$ is the magnitude of the second largest eigenvalue of $\hM$ restricted to the subspace $\mk$ and $d^{(\mk)}$ is a constant factor encoding its degeneracy along with any complex phases.

Using Eq.~\eqref{eq:Ktexpansion}, we see that the SFF $K_{\infty}(t; \mk)$ approaches the linear in $|t|$ RMT behavior after a time
\begin{equation}
    t \simeq \frac{1}{\Delta^{(\mk)}_{\hM}(L)} \equiv \tth^{(\mk)} \, .
\label{eq:tthbound}
\end{equation}
Thus, we have shown that extracting the scaling behavior of $\tth$ with system size $L$ is equivalent, in the large-$q$ limit, to the problem of obtaining the scaling of the ``gap"\footnote{The gap of a bistochastic matrix is traditionally defined to be $1 - \exp\left(-\Delta^{(\mk)}_{\hM}\right)$, which we have approximated as $\Delta^{(\mk)}_{\hM}$ here.} $\Delta^{(\mk)}_{\hM}(L)$ of the Markov circuit $\hM$ (within the subspace $\mk$) with $L$. 
Henceforth, we will restrict our attention to a single (exponentially large) subspace $\mk$ of $\hM$ and suppress the sub/superscript $\mk$ for ease of notation.
To determine the gap $\Delta_{\hM}$, we will now establish a relation between $\hM$ (within a subspace $\mk$) and a quantum Hamiltonian.
We proceed by first observing that $\hM$ corresponds to the \textit{classical} stochastic time evolution of a probability density $\vec{p}(t)$, defined over all product states in the usual $Z$-basis for the spin Hilbert space:  
\begin{equation}
    p_\alpha(t + 1) = \sumal{\beta}{}{\hM_{\alpha\beta}\ p_\beta(t)} \, ,
\label{eq:pevolution}
\end{equation}
where $\hM_{\alpha\beta}$ represents the matrix elements of $\hM$ and the bistochasticity of $\hM$ ensures that the total probability is conserved under time-evolution.
In particular, under the action of each local gate $\hm_{[j, j + \ell - 1]}$ (see Eq.~\eqref{eq:m_gates}), the probability density $\vec{p}$ evolves (with equal probability) to all product states that can be reached via the allowed local moves i.e., moves that are allowed by the constraints imposed on the spin DOFs. 
Now consider starting with a probability density where all the weight is concentrated on a single product state within some subspace $\mk$. Under the stochastic evolution, the probability density will eventually reach a unique equilibrium state, specified by the uniform distribution over all product states in $\mk$ i.e., by the eigenvector of $\hM$ corresponding to the eigenvalue $\Lambda_1 = 1$. Thus, obtaining $\Delta_{\hM}$ is related to obtaining the inverse of the mixing time for this process, which is in general not analytically tractable.  
To derive the gap $\Delta_{\hM}$, we are interested in the stochastic evolution under $\hM$ at time-scales of $\mathcal{O}(1/\Delta_{\hM})$. If $\Delta_{\hM} \rightarrow 0$ in the thermodynamic limit (i.e., if $\hM$ is gapless), we expect that the dynamics under $\hM$ at late times is well approximated by a continuous-time process.\footnote{Note that this approximation does not hold for gapped systems as they relax to their equilibrium distribution on time-scales of $\mathcal{O}(1)$, that are much smaller than the time-scale at which the continuous time description is valid.}
In particular, since $\hM$ corresponds to a stochastic process with local moves occurring independently with equal probability, its late-time behavior should be well approximated by that of a continuous time process composed of the same local moves occurring at equal rates, with the additional requirement of detailed balance.
In other words, the evolution of the probability density $\vec{p}$ should be governed by a Master equation of the form:
\beq
\label{eq:master_eq}
\frac{d p_\alpha (t)}{dt} = \sum_{\beta \neq \alpha } {\left(T_{\alpha \beta} \, p_\beta (t) - T_{\beta \alpha} \, p_{\alpha}(t)\right)} \, ,
\eeq
where $p_\alpha(t)$ is the probability of a classical system occupying state $\alpha$ and $T_{\alpha \beta}$ is the transition rate from state $\beta$ to state $\alpha$, which we specify below. 
Defining $T_{\alpha \alpha} \equiv  - \sumal{\beta \neq \alpha}{}{T_{\beta \alpha}} $, we can rewrite Eq.~\eqref{eq:master_eq} as a matrix equation in terms of the transition matrix $T$,
\beq
\label{eq:master_eq_matrix}
\frac{d \vec{p}(t)}{dt} =  T \, \vec{p}(t) \, ,
\eeq
which ensures that local moves that occur with equal probability in the discrete-time process Eq.~(\ref{eq:pevolution}) occur at equal rates in the continuous-time process Eq.~(\ref{eq:rkfokkerplanck}).
The late time behavior of the stochastic process governed by $\widehat{M}$ is then given by Eq.~\eqref{eq:master_eq_matrix} with 
\beq
\label{eq:rkfokkerplanck}
T = -\Gamma  H \, \implies \dot{\vec{p}}(t) = - \Gamma H \vec{p}(t) \,  ,
\eeq
where $\Gamma$ is an overall positive constant\footnote{As we show in Sec.~\ref{sec:multipole}, $\Gamma$ is a non-universal constant determined by the detailed microscopic properties of the underlying FRQC e.g., it depends on the number of layers $\ell$. However, obtaining its precise value is not important for our purposes.} that sets the rate at which local moves occur and $H$ is defined as
\begin{equation}
    H = \sumal{j}{}{\Pi_{[j, j + \ell - 1]}} \, , \quad \Pi_{[j, j + \ell - 1]} \equiv \mathbb{1} - \hm_{[j, j + \ell - 1]} \, .
\label{eq:HRKdefn}
\end{equation}
The matrix $\Pi_{[j, j + \ell - 1]}$ is a projector and has the form
\begin{align}
\Pi_{[j, j + \ell - 1]} = & \, \bigoplusal{\alpha=1}{\mathcal{D}}{\pi(d_\alpha)} \, , \nn \\ 
\pi(d_\alpha) = & \, \frac{1}{d_\alpha}
\begin{bmatrix}
(d_\alpha -1) & -1 & \dots
\\
-1 & (d_\alpha - 1) & \dots
\\
\vdots & \vdots & \ddots 
\end{bmatrix}_{d_\alpha \times d_\alpha} \, ,
\label{eq:Pi_gates}
\end{align}
thereby ensuring that the transitions taking place in the continuous-time process are identical to those specified by the gates $\hm_{[j, j + \ell - 1]}$. 
In fact, $H$ can be interpreted as a quantum Hamiltonian in the product state basis (in the $Z$-basis) for the spin Hilbert space and has the same symmetries and Krylov subspaces as the stochastic circuit $\hM$. More importantly, $H$ belongs to the class of so-called \textit{RK-Hamiltonians}, where an RK-Hamiltonian is defined as a quantum Hamiltonian that is proportional to the transition matrix $T$ of a discrete classical stochastic process which satisfies detailed balance~\cite{castelnovo2005}. Consequently, the ground state wave function of an RK-Hamiltonian can be interpreted as a classical equilibrium distribution, its low-lying excited states correspond to classical relaxation modes, and its gap coincides with the relaxation time of the corresponding transition matrix. Such Hamiltonians were first studied in the context of quantum dimer models~\cite{rkham} and have subsequently been explored extensively in various settings~\cite{henley1997relaxation,henley2004,castelnovo2005}.

\begin{figure*}[t!]
\begin{tabular}{cc}
\includegraphics[scale = 1]{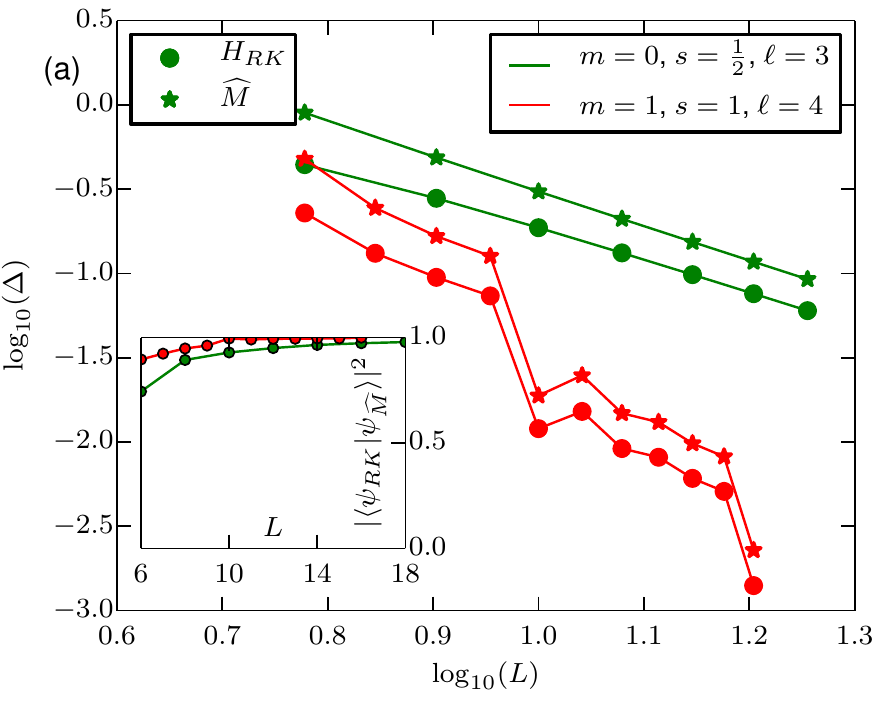}&\includegraphics[scale = 1]{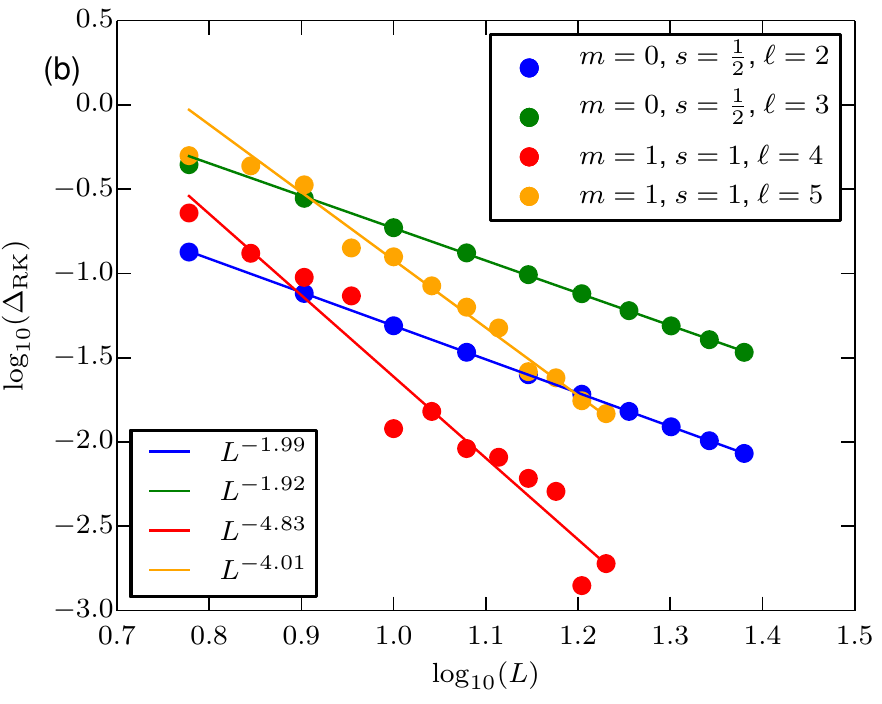} 
\end{tabular}
\caption{(Color online) Scaling of the gaps $\Delta_{\hM}(L)$ and $\Delta_{\mathrm{RK}}(L)$ of $\hM$ and $\hrk$ respectively for a system of spin-$s$ and $\ell$-sized gates with conserved charges $\{Q_j\}$, $j \leq m$. (a) Gaps of $H_{\mathrm{RK}}$ and $\widehat{M}$ for two systems: one with charge conservation and one with dipole conservation. Note that for large system sizes, the gaps are related by a constant factor (see Eq.~(\ref{eq:gapapprox})). The inset shows the overlap of normalized ``first-excited eigenstates" of $\hrk$ ($\ket{\psi_{RK}}$) and $\widehat{M}$ ($\ket{\psi_{\widehat{M}}}$). Note that the overlap approaches 1 with increasing system size, suggesting that the $\ket{\psi_{RK}}$ is an asymptotically exact first-excited state of $\widehat{M}$.  (b) Scaling of the gaps of $H_{\mathrm{RK}}$ with system size for systems with charge (green, blue) and dipole moment (orange, red) conservation. Note that the gaps scale diffusively ($\sim L^{-\beta}$, $\beta \approx 2$) in the presence of only charge conservation whereas they scale subdiffusively ($\sim L^{-\beta}$ with $\beta > 2$) in the presence of dipole moment conservation. All data presented corresponds to the largest symmetry sector/Krylov subspace containing the state $\ket{0\ 0\ \cdots\ 0\ 0}$ for OBC. }
\label{fig:gapscaling}
\end{figure*}

To emphasize the relation between $\widehat{M}$ and an emergent RK-Hamiltonian, we henceforth adopt the notation $H \to \hrk$. In the picture developed above, Eq.~(\ref{eq:rkfokkerplanck}) then has the clear interpretation of an imaginary-time Schr\"{o}dinger evolution under $\hrk$ Eq.~\eqref{eq:HRKdefn}. In effect, the correspondence between $\hM$ and $\hrk$ amounts to a relation between $\Tr_\mk[\hM^t]$ and the partition function of $\hrk$ restricted to $\mk$ at an inverse temperature $\beta = \Gamma t$, namely: $\Tr_\mk[\hM^t] \approx \Tr_\mk[e^{-\Gamma \hrk t}]$. We can therefore approximate the SFF Eq.~(\ref{eq:Kexpand}) at late times as
\begin{equation}
\label{eq:lateK}
    K_\infty(t; \mk) \overset{t \gg 1}{\approx} |t| \textrm{Tr}_{\mk}\left[e^{- \Gamma \hrk t} \right],
\end{equation}
such that the gap of $\hM$ is related to $\Delta_{\mathrm{RK}}(L)$, the gap of the Hamiltonian $\hrk$ (restricted to the subspace $\mk$):
\begin{equation}
    \Delta_{\hM}(L) \approx \Gamma \Delta_{\mathrm{RK}}(L).
\label{eq:gapapprox}
\end{equation} 
Indeed, as we discuss in Sec.~\ref{sec:multipole}, we find numerical evidence that supports Eq.~\eqref{eq:gapapprox} in multipole conserving circuits. 
Further evidence for the correspondence between $\hrk$ and $\hM$ is obtained by studying the system-size dependence of the overlap between the ``first-excited eigenstates" of $\ket{\psi_{\mathrm{RK}}}$ and $\ket{\psi_{\hM}}$ of $\hrk$ and $\hM$ respectively. As shown in the inset of Fig.~\ref{fig:gapscaling}(a), we find that this overlap approaches $1$, suggesting that $\ket{\psi_{\mathrm{RK}}}$ is an asymptotically exact eigenstate of $\hM$ in the thermodynamic limit. We also numerically observe that the overlap does not approach $1$ in the cases when $\hM$ is gapped, further suggesting that correspondence between $\hM$ and $\hrk$ is only valid when $\hM$ is gapless.
The preceding discussion shows that obtaining the gap of $\hrk$ is sufficient for obtaining the scaling of the Thouless time $\tth$. Taken together, Eqs.~\eqref{eq:tthbound} and~\eqref{eq:gapapprox} constitute one of the central results of this paper, whereby a dynamical property of the FRQC is determined by the low-energy, equilibrium behavior of an emergent quantum Hamiltonian. Since we made no reference to the microscopic structure of the underlying circuit, this relationship between $\tth$ and $\Delta_{\mathrm{RK}}$ holds generally for the class of circuits specified in Sec.~\ref{sec:FRQC}.

We now briefly discuss some properties of the Hamiltonians $\hrk$. Denoting the basis set of $\pi(d_\alpha)$ in Eq.~(\ref{eq:Pi_gates}) by $\mB_\alpha$, we obtain
\begin{equation}
    \pi(d_\alpha) = \frac{1}{d_\alpha} \sumal{\mC, \mC' \in \mB_\alpha}{}{\left(\ket{\mC} - \ket{\mC'}\right)\left(\bra{\mC} - \bra{\mC'}\right)},
\label{eq:rkblocks}
\end{equation}
where $\mC$ and $\mC'$ represent $\ell$-site configurations chosen from the basis set $\mB_\alpha$. We can then re-express $\hrk$ as
\begin{align}
    \hrk = & \, \sumal{\langle \mC, \mC'\rangle}{}{w_{\mC, \mC'} \hQ_{\mC, \mC'}},\nn \\
    \hQ_{\mC, \mC'} \equiv & \, \left(\ket{\mC} - \ket{\mC'}\right)\left(\bra{\mC} - \bra{\mC'}\right),
\label{eq:hrkstandardform}
\end{align}
where $\langle \mC, \mC' \rangle$ denotes product states $\mC$ and $\mC'$ that are connected under the action of $\hrk$ and the weights $w_{\mC, \mC'} \geq 0$ are defined in accordance with Eq.~\eqref{eq:rkblocks}. From this expression, it is clear that $\hrk$ is a positive semidefinite Hamiltonian and the zero-energy ground state wave function within a subspace $\mk$ is given by 
\begin{equation}
    \ket{G^{(\mk)}_\mathrm{RK}} = \frac{1}{\sqrt{\mZ}}\sumal{\mC \in \mk}{}{\ket{\mC}} \, ,
\label{eq:RKGS}
\end{equation}
where $\mC$ runs over all the product states in the subspace $\mk$ and $\mZ$ is a normalization factor. 
We remark that, up to an overall normalization factor, $\grk$ can be interpreted as the equilibrium probability distribution of the stochastic process described by Eqs.~\eqref{eq:pevolution} or~\eqref{eq:rkfokkerplanck}.

Before proceeding to focus on multipole conserving circuits, we briefly discuss some important aspects of the RK-Hamiltonian $\hrk$ which illustrate the potential benefits of the mapping developed in this section.    
First, $\hrk$ is a frustration-free Hamiltonian, i.e., the ground state $\ket{G_{\textrm{RK}}}$ is the ground state of each of the terms $\Pi_{[j, j + \ell - 1]}$ as can be seen using Eqs.~\eqref{eq:hrkstandardform} and~\eqref{eq:RKGS}, since $\hQ_{\mC, \mC'}\ket{G^{(\mk)}_{RK}} = 0$ for any $\mC$ and $\mC'$.
This fact enables the use of well-known methods for bounding the spectral gap of frustration-free Hamiltonians~\cite{knabe1988energy, gosset2016local, lemm2019spectral}, which in turn allow us to place a constraint on the scaling exponent of $\tth$ in FRQC with constraints defined in Eq.~\eqref{eq:wdef} in the large $q$ limit:
\beq
\label{eq:genlimit}
\tth \sim L^\alpha \, , \quad
\begin{cases} 
\alpha = 0 \textit{ or } \alpha \geq 2, & \text{PBC} \\
\alpha = 0 \textit{ or }\alpha \geq 3/2, & \text{OBC}
\end{cases}
\eeq
We note that for circuits \textit{without} any conserved quantities, $\tth$ has been shown to scale as $\log L$ for certain Floquet models~\cite{cdc2, kos2018}. However, as evidenced by Eq.~\eqref{eq:genlimit}, for circuits of the form Eq.~\eqref{eq:wdef} (including the circuit discussed in Ref.~\cite{cdc1}), such scaling of $\tth$ is suppressed in the $q \to \infty$ limit and we instead find that $\tth$ can scale as an $\mathcal{O}(1)$ number in FRQCs in this limit.

Secondly, by virtue of the connection to classical Master equations, shown in Eq.~(\ref{eq:rkfokkerplanck}), $\hrk$ is an example of a \textit{stoquastic} Hamiltonian, which can be efficiently studied using Quantum Monte Carlo techniques~\cite{castelnovo2005, bravyi2015monte}. 
We thus expect that the same techniques can be exploited to efficiently study the late-time features of the SFF in a variety of settings at large-$q$.
Moreover, in the context of spectral graph theory, any Hamiltonian of the form Eq.~(\ref{eq:hrkstandardform}) restricted to a subspace $\mk$ exactly corresponds to the Laplacian~\cite{chung1997spectral} of an undirected graph $\mG$, formed by the set of vertices $\{\mC\}$ within $\mk$ and by edges with weights $w_{\mC, \mC'}$ between the vertices $\mC$ and $\mC'$.
The gap $\Delta_{\mathrm{RK}}$ then corresponds to the gap of the Laplacian of the graph $\mathcal{G}$, which is closely related to the connectivity of $\mG$. In particular, the existence of bottlenecks in $\mG$, as detected by the Cheeger constant, results in a smaller gap of the Laplacian (and hence, in a larger $\tth$).
This establishes a clear connection between the nature of transport in the presence of constraints and the connectivity of the Hilbert space under those constraints.
Finally, we note that earlier work has also discussed a relation between the Thouless time $\tth$ of a charge-conserving FRQC and and the spectral gap of a U(1) invariant classical bistochastic circuit: Ref.~\cite{friedman2019} constitutes a particular case of the results obtained in this paper, being derived in the large-$q$ limit, while Ref.~\cite{roy2020random} invokes the random phase approximation in a long-range interacting model at finite-$q$. While both of these works were restricted to specific realizations of U(1) invariant systems, the relations between $K_\infty$, $\hM$, and $\hrk$ obtained in this section apply far more generally to the large class of circuits with arbitrary symmetries or constraints discussed in Sec.~\ref{sec:FRQC}.
%
%
%
%

\section{Examples from Multipole Conserving Circuits}
\label{sec:multipole}

In this section, we move our attention to FRQCs with conserved higher moments and provide explicit examples of the mapping established in Sec.~\ref{sec:mapping}. Specifically, we consider circuits which conserve all moments of charge up to the $m^{th}$ highest moment, where the $m^{th}$ multipole moment is defined as
\begin{equation}
    \hQ_m = \begin{cases} 
    \sumal{x = 1}{L}{x^m S^z_x} \, , & \text{OBC} \\
    \exp\left(2\pi i \sumal{x = 1}{L}{(x/L)^m S^z_x} \right) \, , & \text{PBC}
    \end{cases}
\label{eq:multipole}
\end{equation}
where $m = 0 \, (1)$ corresponds to the charge (dipole) conserving case. Where necessary, we will use the labels $\{Q_m\}$ to denote the $m^{th}$ multipole moment quantum number. 
We start by reviewing the charge conserving FRQC (see Sec.~\ref{sec:FRQC}), which was previously discussed in Ref.~\cite{friedman2019}.
Following the general discussion in Secs.~\ref{sec:FRQC} and~\ref{sec:mapping}, the stochastic circuit $\hM$ for a spin-1/2 U(1) charge conserving circuit with gate size $\ell = 2$ (for PBC) is given by
\begin{align}
    \hM = & \, \bigotimesal{j\ \textrm{odd}}{}{\hm_{[j, j + 1]}}\bigotimesal{j\ \textrm{even}}{}{\hm_{[j, j+1]}} \, , \nn \\
    \hm_{[j, j+1]} = & \,
    \begin{bmatrix}
        1 & 0 & 0 & 0 \\
        0 & 1/2 & 1/2 & 0 \\
        0 & 1/2 & 1/2 & 0 \\
        0 & 0 & 0 & 1
    \end{bmatrix} \, ,
\label{eq:chargecons}
\end{align}
where the local 2-site gate $\hm_{[j,j +1]}$ is written in the (ordered) basis $\{ \ket{\uparrow \uparrow}, \ket{\uparrow \downarrow}, \ket{\downarrow \uparrow}, \ket{\downarrow \downarrow} \}$. 
Using Eqs.~(\ref{eq:Pi_gates}) and~(\ref{eq:chargecons}), we find that $\Pi_{[j, j+1]}$ maps onto a ferromagnetic spin-1/2 Heisenberg term
\begin{eqnarray}
    \Pi_{[j, j + 1]} &=& 
    \begin{bmatrix}
        0 & 0 & 0 & 0 \\
        0 & 1/2 & -1/2 & 0 \\
        0 & -1/2 & 1/2 & 0 \\
        0 & 0 & 0 & 0
    \end{bmatrix} \nn \\
    &=& \frac{1}{2}\left(\ket{\uparrow \downarrow} - \ket{\downarrow\uparrow}\right)\left(\bra{\uparrow \downarrow} - \bra{\downarrow \uparrow}\right)_{j,j+1} \nn \\
    &=& \frac{1}{4}\left(1 - \vec{\sigma}_j\cdot \vec{\sigma}_{j+1}\right) \, ,
\label{eq:heisenberg}
\end{eqnarray} 
where $\Pi_{[j,j +1]}$ is written in the (ordered) basis $\{ \ket{\uparrow \uparrow}, \ket{\uparrow \downarrow}, \ket{\downarrow \uparrow}, \ket{\downarrow \downarrow} \}$ and $\vec{\sigma} = (\sigma^x, \sigma^y, \sigma^z)$, with $\sigma^i$ the usual Pauli matrices.
For the charge conserving FRQC (with $\ell = 2$), we hence find that $\hrk$ is the Bethe-Ansatz integrable ferromagnetic Heisenberg model, whose integrability was exploited in Ref.~\cite{friedman2019} to study the late-time behavior of the SFF for this circuit. 

According to Eq.~(\ref{eq:RKGS}), the unique ground state within any charge sector $Q_0$ is the equal amplitude superposition of all product states within that symmetry sector. Indeed, such a state belongs to the SU(2) multiplet of the spin-polarized ferromagnetic state with total spin $Q_0 = L$. 
Moreover, the low-energy excitations above the ferromagnetic state in the Heisenberg model of Eq.~(\ref{eq:heisenberg}) are exactly known to be spin waves with dispersion $\epsilon(k) = 2 \sin^2\left( k/2 \right)$.
As a consequence of the SU(2) symmetry of the Heisenberg model, the lowest energy excited state within each symmetry sector belongs to the multiplet of spin-wave states with total spin $Q_0 = L - 1$; the gap of $\hrk$ in any $Q_0 \neq L$ sector is then given by 
\beq
\Delta_{\mathrm{RK}}(L) = \epsilon\left(k = \frac{2\pi}{L}\right) \approx \frac{\pi^2}{2L^2} \, ,
\eeq
which is the energy corresponding to the lowest non-zero momentum spin-wave.  
Using Eqs.~(\ref{eq:tthbound}) and~\eqref{eq:gapapprox}, we find that the Thouless time in any quantum number sector in an FRQC with U(1) charge conservation scales diffusively with system size i.e., $\tth \sim L^2$.
For charge conserving systems with higher spins or larger gate sizes $\ell$, $\hrk$ is no longer integrable in general, but, as shown in Fig.~\ref{fig:gapscaling}(b), we numerically observe the same diffusive scaling $\Delta_{\hM}(L) \sim \Delta_{\mathrm{RK}} \sim L^{-2}$ for the systems we studied. In fact, we find that the gaps are identical for spin-1/2 and spin-1 systems with gate size $\ell = 2$ even though the rest of the spectrum is different, strongly suggesting a universal origin of the scaling.

As evidenced through the above example, the correspondence between the FRQC and the RK-Hamiltonian unveils a curious feature of the large-$q$ limit. While the original FRQC only has U(1) symmetry, after Haar averaging and taking $q\to\infty$, $K_\infty(t;\mk)$---related to $\hrk$ through Eq.~\eqref{eq:lateK}---exhibits an enlarged SU(2) invariance in the spin DOFs. Indeed, we expect this enlarged symmetry to be a generic feature in the large-$q$ limit, since RK-points typically exhibit enhanced symmetries, although not necessarily SU(2)~\cite{fradkin2004RK, ardonne2004, Moessner2011}. On the other hand, to our knowledge, the emergent integrability in the above example is \textit{not} generic and is specific to the spin-1/2 system with 2-site gates.
We now turn our attention to systems which conserve the dipole moment $\hQ_1$ in addition to the charge $\hQ_0$, for which the nature of low-energy excitations above the ground state of the corresponding RK-Hamiltonian $\hrk$ is not immediately apparent. 
An additional feature in such systems is the fragmentation of the Hilbert space of the FRQC~\cite{sala2020ergodicity, khemani2020}, which leads to the formation of exponentially many Krylov subspaces (see Eq.~(\ref{eq:Wkryl})). 
Hilbert space fragmentation is typically classified into two types: strong or weak, where the size of the largest Krylov subspace is respectively a zero or non-zero fraction of the total Hilbert space dimension within a given quantum number sector in the thermodynamic limit.
Refs.~\cite{sala2020ergodicity, khemani2020} numerically observed that spin-1 and spin-1/2 dipole conserving systems with the minimal gate sizes $\ell = \ell_{min} = 3$ and $\ell = \ell_{min} = 4$ respectively show strong fragmentation whereas the inclusion of moves requiring larger gate sizes leads to weak fragmentation.
Furthermore, Ref.~\cite{morningstar2020} found that the nature of fragmentation can vary even for a given gate size depending on the quantum number sector. 
Generically, however, experimentally relevant multipole conserving systems are expected to show weak fragmentation.

In strongly fragmented systems, the ratio between the dimension of the largest Krylov subspace within a symmetry sector and the size of that symmetry sector exponentially decays to zero in the thermodynamic limit. As a consequence, typical initial states do not thermalize~\cite{sala2020ergodicity, khemani2020}, although certain initial states do thermalize with respect to smaller Krylov subspaces~\cite{moudgalya2019quantum, moudgalya2019krylov}. In contrast, for weakly fragmented systems there always exists a dominant Krylov subspace $\mk$ within a given quantum number sector, such that its size asymptotically approaches that of the symmetry sector in the thermodynamic limit. Due to this, typical eigenstates within a quantum number sector carry non-zero weight in the dominant Krylov subspace of that symmetry sector and look thermal. As a consequence, frozen configurations, despite being exponential in number, are expected to have a negligible effect on $\tth$ in a weakly fragmented system.
Since our interest in this work is the behavior of \textit{generic} multipole conserving systems, we focus only on thermalizing weakly fragmented systems here.
Hence, we will study the SFF, the scaling of the Thouless time $\tth$ and the gaps $\Delta_{\hM}(L)$ and $\Delta_{\mathrm{RK}}(L)$ all restricted to the dominant Krylov subspace $\mk$ within a specified quantum number sector.

In Fig.~\ref{fig:gapscaling}, we show the scaling of the gaps $\Delta_{\hM}(L)$ and $\Delta_{\mathrm{RK}}(L)$ for spin-1/2 and spin-1 dipole conserving systems for several gate sizes $\ell > \ell_{min}$.
Fig.~\ref{fig:gapscaling}(a) shows that the numerics are in good agreement with Eq.~\eqref{eq:gapapprox} i.e., they support the correspondence between the stochastic circuit $\hM$ and the emergent RK-Hamiltonian $\hrk$ developed in Sec.~\ref{sec:mapping}. In principle, we can also extract the microscopic constant $\Gamma$ for a specific circuit by comparing the gaps for the corresponding $\hM$ and $\hrk$. 
Furthermore, as evident from the numerics shown in Fig.~\ref{fig:gapscaling}(b), we find that $\Delta_{\mathrm{RK}}(L)$ scales as $\sim L^{-\beta}$ with $\beta > 2$; thus, the Thouless time scales \textit{subdiffusively} $\tth \sim L^\beta \, (\beta > 2)$ for system sizes accessible to exact diagonalization. Importantly, this subdiffusive scaling appears to be a \textit{generic} feature of weakly fragmented, dipole conserving systems and does not show a strong dependence on the microscopic details ($\ell$ or $s$) of the circuit. This mirrors the behavior of systems with only charge conservation (Fig.~\ref{fig:gapscaling}(a)), which show diffusive scaling of $\tth$ independent of microscopic details.

Due to the longer gate sizes required for systems conserving even higher multipole moments $Q_{m\geq 2}$ (e.g., $\ell_{min} = 8$ for spin-1/2 quadrupole conserving systems, with longer gates likely needed for weak fragmentation), we are unable to eliminate finite-size effects in such cases. 
Nevertheless, the numerics suggest a universality in the scaling of the gaps of charge and dipole conserving RK-Hamiltonians $\hrk$ i.e., $\Delta_{\mathrm{RK}}(L) \sim L^{-2}$ for charge conserving systems and $\Delta_{\mathrm{RK}}(L) \sim L^{-4}$ for dipole conserving systems, regardless of the ultraviolet details of the respective Hamiltonians. The appearance of this universality suggests the existence of a universal field theoretic description which effectively captures the low-energy behavior of generic multipole conserving RK-Hamiltonians, such as the scaling of their gap. The derivation of these universal effective field theories will be the subject of the next section.
%
%
\section{Continuum Limit for multipole conserving systems}
\label{sec:continuum}
As discussed in the previous section, the ground state of an RK-Hamiltonian is well-known as being the equal-weight superposition of all states in the corresponding Hilbert space. However, understanding the scaling of the gap $\Delta_{\mathrm{RK}}$ requires knowledge of low-lying states above the GS, which are generically not known exactly. 
Nevertheless, motivated by our numerical observation of a universal scaling of $\Delta_{\mathrm{RK}}$ with $L$ for generic charge and (weakly fragmented) dipole conserving RK-Hamiltonians, we derive continuum field theories for multipole conserving systems through a coarse-graining procedure, detailed in Appendix~\ref{app:continuumhamil}. We find that the resultant continuum field theories accurately capture the ground state and low-energy excitations of the corresponding RK-Hamiltonians, therefore providing an analytic route to understanding the scaling of $\tth$ in the underlying FRQC. 

Throughout this section, we will only consider OBC. We denote the number of spins as $N$ and the system size as $L = N \Delta x$, where $\Delta x$ is the lattice spacing. The continuum limit then corresponds to taking the limits $N \to \infty$ and $\Delta x \to 0$ simultaneously, while keeping $L$ fixed. 
For systems which conserve all moments of charge up to the $m^{th}$ moment (or, $m^{th}$ moment conserving systems), we focus our attention on the quantum number sector $\ms= \{Q_0=0,\ Q_1 = 0,\  \cdots,\ Q_m = 0\}$.
As discussed in Sec.~\ref{sec:multipole}, for weakly fragmented systems there exists a dominant Krylov subspace within each symmetry sector, such that the size of that subspace asymptotically approaches the size of the full symmetry sector as the gate-size $\ell$ increases.
Since taking the continuum limit involves coarse-graining and thus effectively taking the gate-size $\ell \rightarrow \infty$, we can neglect the effect of fragmentation in systems with dipole and higher moment conservation, and expect that our analysis holds as long as the sectors we study do not exhibit strong fragmentation.

\subsection{Generalized height fields}
\label{sec:genheight}

\begin{figure}[t!]
\centering
\includegraphics[scale = 0.6]{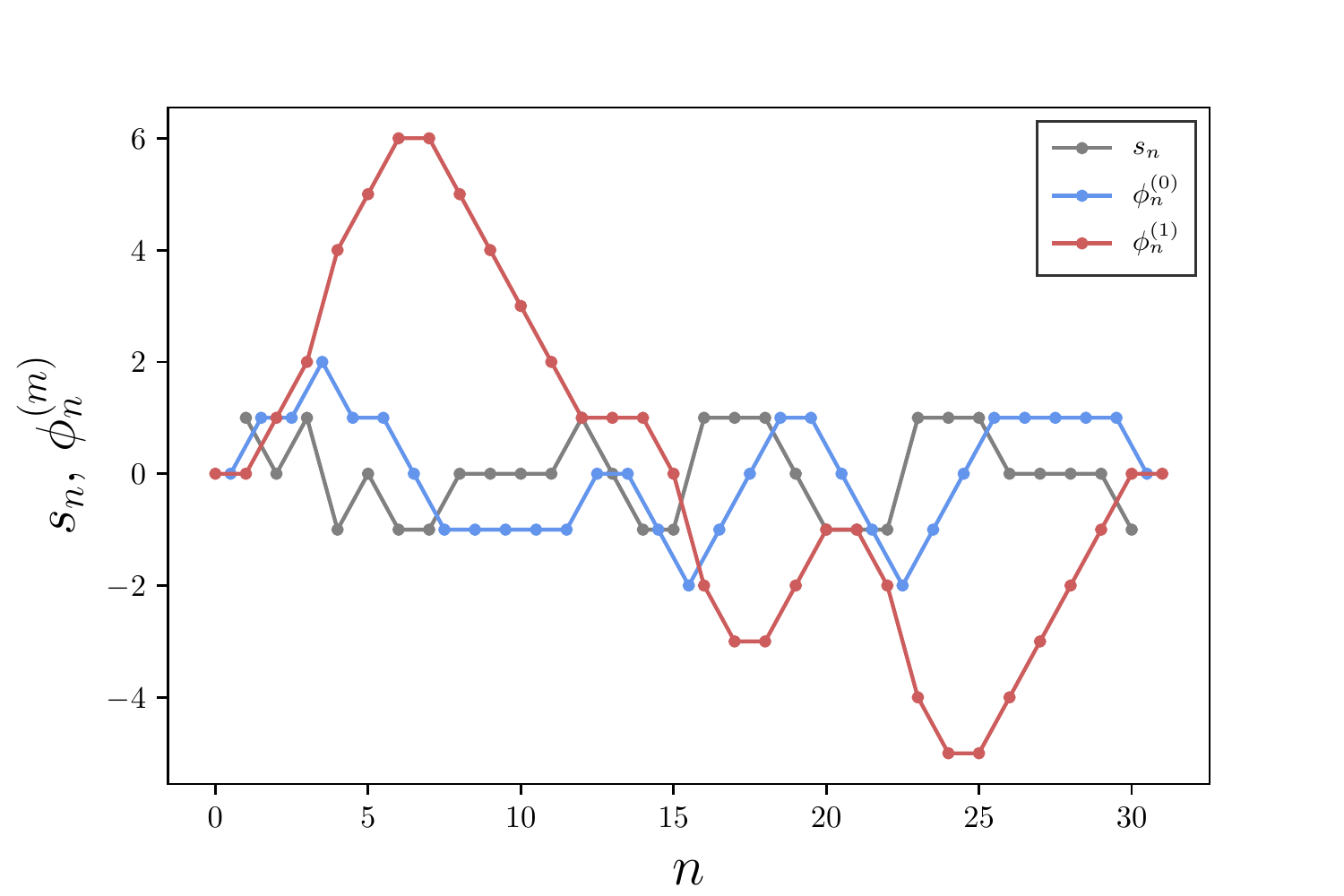}
\caption{(Color online) The generalized height variables of an example spin state with $Q_0=0$ and $Q_1=0$ for $L=30$. Note that the end points $\phi^{(2)}_0$ and $\phi^{(2)}_{N + 1}$ are fixed to be zero.}
\label{fig:height_field_example} 
\end{figure}

In order to take the continuum limit, we first need to introduce ``generalized" height fields, in analogy with familiar height fields in the quantum dimer context~\cite{Moessner2011}. When taking the continuum limit of the ground state wave function and $\hrk$, a crucial issue is the restriction to the symmetry sector $\ms$; this restriction imposes a \textit{global} constraint on the spin DOFs $\{s_n\}$, thereby resulting in a non-local action for system. It is to circumvent precisely this issue, while preserving locality, that we work in terms of generalized height variables $\{\phim\}$ for systems with $m^{th}$ moment conservation. In terms of these variables, the conservation of higher moments $\{Q_m\}$ are expressed as local \textit{boundary constraints} on the height fields and their derivatives, as opposed to a global constraint on the spin DOFs.  
We first illustrate this construction in terms of height variables for systems with charge conservation. The height DOFs $\{\phi^{(0)}_{n + \frac{1}{2}}\}$ are defined on the links of the one-dimensional chain as
\begin{equation}
    \phi^{(0)}_{n + \frac{1}{2}} -  \phi^{(0)}_{n - \frac{1}{2}} = s_n\;\;\;\textrm{or}\;\;\; \phi^{(0)}_{n + \half} = \phi^{(0)}_{\half} + \sumal{j = 1}{n}{s_n} \, ,
\label{eq:heightdefn}
\end{equation}
which immediately suggests that the total charge $Q_0$ (see Eq.~\eqref{eq:multipole}) is given by the flux of the height variable through the system:
\begin{equation}
    Q_0 = \phi^{(0)}_{N + \half} - \phi^{(0)}_{\half} \, .
\label{eq:totalcharge}
\end{equation}
Since Eqs.~(\ref{eq:heightdefn}) and (\ref{eq:totalcharge}) are invariant under an overall constant shift of the height variables ($\phi_{n+1/2}\to \phi_{n+1/2} + c$), we can choose $\phi^{(0)}_{\half} = 0$ without loss of generality. 
Thus, restricting to a given charge sector corresponds to imposing constraints on the boundary height variables once we forgo the spin language for the height representation.
Furthermore, as a consequence of Eq.~(\ref{eq:heightdefn}), any charge conserving process involving $\ell$  spins $\{s_n,\ \cdots\ , s_{n + \ell - 1}\}$ involves $(\ell - 1)$ height variables $\{\phi^{(0)}_{n + \half},\ \cdots\ \phi^{(0)}_{n + \ell - \frac{3}{2}}\}$. So, the mapping from the spin DOFs to the height variables preserves the locality of the Hamiltonian. 
We now generalize the height representation to systems with $m^{th}$ moment conservation. For a system with all moments up to the $m^{th}$ moment conserved, we recursively define the $m^{th}$ ``generalized" height variable $\phim$, that lives on links (sites) if $m$ is even (odd), as
\begin{align}
     &\twopartdef{\phim_{n + \half} - \phim_{n - \half} = \phi^{(m-1)}_{n}}{m \in 2\mathbb{Z}}{\phim_{n + 1} - \phim_{n} = \phi^{(m-1)}_{n + \half}}{m \in 2\mathbb{Z} + 1},
\label{eq:recursiveheight}
\end{align}
with $\{\phi^{(0)}_{n + \half}\}$ defined in Eq.~(\ref{eq:heightdefn}). 
An example of a charge configuration in a dipole conserving ($m=1$) system, expressed in terms of height variables, is depicted in Fig.~\ref{fig:height_field_example}. 
As mentioned earlier, the key advantage of forgoing the spin language is that the total $m^{th}$ multipole moments can be expressed in terms of boundary height variables and their ``derivatives", an observation that will prove essential when taking the continuum limit.  
For instance, the total dipole moment can be expressed as
\begin{align}
    Q_1 =& \sumal{j = 1}{N}{j s_j} = N Q_0 - \sumal{j = 1}{N-1}{\sumal{k = 1}{j}{s_k}} \nn \\
    =&  N\left(\phi^{(1)}_{N +1} - \phi^{(1)}_{N}\right) - \left(\phi^{(1)}_{N} - \phi^{(1)}_{0}\right),
\label{eq:dipoleop}
\end{align}
where we have used Eqs.~(\ref{eq:totalcharge}) and (\ref{eq:recursiveheight}).
Similarly to the charge conserving case, locality is also preserved when going to the generalized height representation. This can be seen from Eq.~(\ref{eq:recursiveheight}), since any $m^{th}$ multipole conserving process involving $\ell$ spins $\{s_n,\ \cdots\ , s_{n + \ell - 1}\}$ involves $(\ell - m + 1)$ generalized height variables $\{\phim_{n + \half} \}$.
We now take the continuum limit by defining generalized height fields through an appropriate rescaling of the height variables by the lattice spacing $\Delta x$ i.e.,
\begin{equation}
    \rho(x) = \frac{s_n}{\dx},\;\;\;\phim(x) = \phim_{n}(\dx)^m \, ,
\label{eq:heightfields}
\end{equation}
where $x = n\Delta x$.
Here, $\rho(x)$ can be interpreted as the charge density and, as we will see, the height fields $\phim(x)$ are related to the multipolar densities. 
In terms of the height fields, Eq.~(\ref{eq:heightdefn}) in the continuum becomes
\begin{equation}
    \partial_x \phi^{(0)}(x) = \rho(x) \, .
\label{eq:heightgauss}
\end{equation}
Note that Eq.~(\ref{eq:heightgauss}) closely resembles a Gauss' Law.
Similarly to Eq.~(\ref{eq:totalcharge}), the total charge $Q_0$ in the continuum is expressed in terms of the height fields as
\begin{equation}
    Q_0 = \int_0^L{dx\ \rho(x)} = \phi^{(0)}(L) - \phi^{(0)}(0) \, .
\label{eq:Q0cont}
\end{equation}

The preceding discussion illustrates how the global charge constraint on $\rho(x)$ is re-expressed as a \textit{local} boundary constraint on $\phi^{(0)}(x)$, clarifying why the latter is physically more appropriate as the field variable for charge conserving systems. 
Similarly, using Eq.~(\ref{eq:heightfields}), we can express Eq.~(\ref{eq:recursiveheight}) in terms of the generalized height fields as
\begin{equation}
    \partial_x \phim(x) = \phi^{(m-1)}(x)\;\;\textrm{or}\;\;\partial_x^{m+1}\phim(x) = \rho(x) \, .
\label{eq:generalizedheight}
\end{equation}
In the continuum, the conservation of the $m^{th}$ and all lower moments then amounts to fixing the left and right boundary constraints on $\phim(x)$ and its derivatives $\partial_x^n \phim(x)$ for all $n \leq m$.
For instance, the total dipole moment and charge can be expressed in terms of boundary constraints on $\phi^{(1)}(x)$ and $\partial_x \phi^{(1)}(x)$ as
\begin{align}
    Q_0 =& \, \partial_x \phi^{(1)}(L) - \partial_x \phi^{(1)}(0) \, , \nn \\
    Q_1 =& \, \int_0^L{dx\ x\ \rho(x)} \nn \\
    =& \, \left.x\ \phi^{(0)}(x)\right|_0^L - \int_0^L{dx\ \phi^{(0)}(x)}\nn \\
    =& \, L\ \partial_x\phi^{(1)}(L) - (\phi^{(1)}(L) - \phi^{(1)}(0)) \, , \nn \\
\label{eq:Q1cont}
\end{align}
where we have invoked Eqs.~(\ref{eq:heightgauss}) and (\ref{eq:generalizedheight}).  
In fact, it is straightforward to show that the $n^{th}$ multipole moment can be expressed in terms of the $m^{th}$ height field as ($m \geq n$)
\begin{align}
    Q_n =& \, \int_0^L{dx\ x^n\ \rho(x)} \nn \\
    = & \, \int_0^L{dx\ x^n\ \partial_x^{m+1}\phim(x)} \nn \\
    =& \, n!\sumal{j = 0}{n}{\frac{(-1)^j}{j!} \left(L^{j} \partial_x^{m-n + j}\phim(L) - \delta_{j,0} \partial_x^{m -n}\phim(0)\right)} \,. \nn \\
\label{eq:nthmoment}
\end{align}
The multipole moments $Q_n$ in Eq.~(\ref{eq:nthmoment}) are invariant under a polynomial shift of the height fields $\phim$, under which Eq.~(\ref{eq:generalizedheight}) is still satisfied: 
\begin{equation}
    \phim(x) \rightarrow \phim(x) + P^{(m)}(x), 
\label{eq:polynomialshift}
\end{equation}
where $P^{(m)}(x)$ is an arbitrary polynomial of degree $\leq m$; Eq.~(\ref{eq:polynomialshift}) can thus be used to set $\partial_x^{n}\phim(0) = 0$ for all $n \leq m$ without loss of generality, so that 
\beq
Q_n = n! \,  \sumal{j = 0}{n}{\frac{(-1)^j L^j}{j!}  \partial_x^{m-n + j}\phim(L) } \,.
\eeq
In the sector of primary interest, $Q_n = 0$ for all $n \leq m$, these boundary constraints further simplify to
\beq
\label{eq:bcspec}
\partial_x^n\phim(L) = 0 \, \forall \, n \leq m \, .
\eeq
For OBC, we need to further supplement these boundary constraints, which fix the symmetry sectors, with \textit{physical} boundary conditions, which ensure that no multipole currents flow through the boundaries. As discussed in Ref.~\cite{gromovhydro}, the fundamental hydrodynamic quantities for multipole conserving systems are the \textit{charge density} $\rho(x)$ and the \textit{multipole current} $J^{(m)}$, from which one can infer the conventional charge current; however, it is the multipole current that is fundamental and is related to the charge density as
\beq
J^{(m)}(x) \sim \partial_x^{m+1} \rho(x) \, ,
\eeq
for systems which conserve all moments up to the $m^{th}$ highest moment, which is the generalization of Fick's law to multipole conserving systems~\cite{feldmeier2020}.
The physical requirement that no multipole current flows through the boundaries, phrased in terms of the height fields, can be stated as
\beq
\label{eq:physbc}
\partial_x^{2(m+1)} \phim(0) = \partial_x^{2(m+1)} \phim(L) = 0.
\eeq

\subsection{Ground state}
\label{sec:GS}
Before we obtain the continuum limit of the Hamiltonian $\hrk$, we express the ground state wave function of an $m^{th}$ moment conserving system, discussed in Sec.~\ref{sec:multipole}, in terms of the height field $\phim(x)$.
Recall that the GS Eq.~\eqref{eq:RKGS} is the equal-weight superposition of all allowed basis states i.e., all possible height variable configurations that satisfy the boundary constraints, which fix the quantum number sectors.
Treating the spin-$s$ DOFs as ``random variables" that assume integer or half-integer values in $[-s, s]$, under coarse-graining the distribution of the spins flows to a Gaussian as a direct consequence of the central limit theorem~\cite{majumdar2005brownian}. 
The variance of the resulting coarse-grained DOFs then scales as $\sigma^2 = \dx/\kappa$ where $\dx$ is the lattice spacing and $\kappa$ is a parameter chosen such that microscopic correlation functions are accurately reproduced at long distances; effectively, one can think of $\kappa$ as the coarse-graining length scale.
After coarse-graining, the wave functional $\Phi^{(m)}_0[\rho(x)]$ corresponding to a charge density profile $\rho(x)$ is thus simply given by a Gaussian, albeit subject to global constraints specified by the conserved quantities (see Appendix~\ref{app:continuumwf} for details):
\begin{equation}\label{eq:gsspins}
\Phi^{(m)}_0 [\rho(x)] = \frac{1}{\sqrt{\mathcal{Z}}} 
\, e^{-\frac{\kappa}{2}\int_0^L{dx\  (\rho(x))^2}} \; \times \;\mathcal{G}[\rho(x)]
\end{equation}
where $\mathcal{Z}$ is a normalization constant and $\mathcal{G}[\rho(x)]$ enforces the global symmetry constraints; namely, it fixes the quantum number sector of interest. For instance, for a dipole conserving system in the $\{Q_0,Q_1\}$ sector, 
\beq
\mathcal{G}[\rho(x)] = \delta \left( \int \rho(x) - Q_0 \right) \delta \left( \int x \rho(x) - Q_1 \right) \, .
\eeq

To circumvent the global constraint in Eq.~(\ref{eq:gsspins}), it is convenient to work in terms of the $m^{th}$ height fields for $m^{th}$ multipole conserving systems---as discussed in Sec.~\ref{sec:genheight}, in this language, the quantum number sectors are instead expressed as local boundary constraints. 
More explicitly, Eq.~(\ref{eq:generalizedheight}) allows us to express the wave functional Eq.~(\ref{eq:gsspins}) in terms of the height field $\phim(x)$ as
\begin{equation}
\label{eq:gsheight}
\Phi^{(m)}_0 [\phim(x)] = \frac{1}{\sqrt{\mathcal{Z}}} 
\, e^{-\frac{\kappa}{2}\int_0^L{dx\ (\partial_x^{m+1} \phim(x))^2}} \; \mathcal{B}[\phim(x)] \, ,
\end{equation}
where the global constraints encoded in $\mathcal{G}[\rho(x)]$ are replaced with local boundary constraints $\mathcal{B}[\phim(x)]$. These constraints are imposed by $\delta$-functions that fix the boundary constraints on the height fields, corresponding to the quantum number sector of interest (see Eq.~\eqref{eq:nthmoment}). For the sector with $Q_n = 0 \, \forall \, n\leq m$
\beq
\mathcal{B}[\phim(x)] = \prod_{n=0}^m \delta\left(\partial_x^n \phim(L) \right) \, ,
\eeq
which follows from Eq.~\eqref{eq:bcspec}. Note that we also need to impose the physical boundary conditions Eq.~\eqref{eq:physbc} on the generalized height fields.
Recall that in the discrete setting, the GS is an equal weight superposition of allowed configurations, while taking the continuum limit introduces Gaussian weights into the GS due to coarse-graining. Concurrently, the corresponding continuum Hamiltonian will no longer be of the form Eq.~\eqref{eq:hrkstandardform} but instead belongs to the class of ``SMF decomposable" Hamiltonians,\footnote{Real, symmetric, and irreducible matrices which admit a Stochastic Matrix Form (SMF) decomposition were found to be in 1-to-1 correspondence with classical stochastic systems described by a master equation in Ref.~\cite{castelnovo2005}} that are related to classical master equations and include the RK-Hamiltonians Eq.~\eqref{eq:hrkstandardform} as a subclass~\cite{castelnovo2005}. This correspondence will prove useful in deriving the continuum expression for the RK-Hamiltonian, which we discuss in Sec.~\ref{sec:hamiltonian} (see also Appendix~\ref{app:smf}).

To close this discussion, we note that expressions of the form Eq.~\eqref{eq:gsheight} have previously been derived for the continuum limit of ground states of RK-Hamiltonians using various methods, albeit never in the context of multipole conserving systems. 
For instance, the exponent in Eq.~(\ref{eq:gsheight}) can be interpreted as the free energy functional corresponding to a configuration of the height field $\phim(x)$, as is typically done in the context of RK points in dimer models~\cite{henley1997relaxation, Moessner2011}.
Alternately, the expression Eq.~(\ref{eq:gsheight}) can also be derived using the path-integral formulation of Brownian motion: here, one interprets $x$ as a time coordinate, $\rho(x)$ in Eq.~(\ref{eq:generalizedheight}) as white-noise, and $\phim(x)$ as a trajectory under the ``Langevin dynamics" described by Eq.~(\ref{eq:generalizedheight})~\cite{henley1997relaxation, Chen_2017a, majumdar2005brownian}.
\subsection{Hamiltonian and dispersion relation}
\label{sec:hamiltonian}

Having obtained the continuum expression for the ground state wave functional, we now identify the corresponding expression for the coarse-grained RK-Hamiltonian $\hrk$.
As discussed in the previous section, the coarse-grained wave functional Eq.~\eqref{eq:gsheight} is the ground state of a multipole conserving RK-Hamiltonian, which belongs to the generalized class of frustration-free positive-definite RK-like Hamiltonians discussed in Ref.~\cite{castelnovo2005}.
In Appendix~\ref{app:continuumhamil}, we discuss two distinct approaches for deriving the continuum limit of $\hrk$: the first approach involves an appropriate choice of regulators, which allows us to explicitly obtain the continuum parent Hamiltonian corresponding to Eq.~\eqref{eq:gsheight}. 
The second, more commonly employed approach~\cite{henley1997relaxation, ardonne2004, Moessner2011, Chen_2017a, Chen_2017b} exploits the relationship between $\hrk$ and classical master equations discussed in Sec.~\ref{sec:mapping} (see Eq.~\eqref{eq:rkfokkerplanck}). In summary, this approach proceeds by identifying the classical process corresponding to $\hrk$ which equilibrates to a Gaussian distribution of height fields, as given by Eq.~(\ref{eq:gsheight}). As we show in Appendix~\ref{app:continuumhamilFP}, this classical process describes the Langevin dynamics of the generalized height fields under damping. The continuum expression for $\hrk$ can then be derived via the Fokker-Planck equation for the probability functionals of the generalized height fields.

Both approaches lead to the same continuum expression for $\hrk$, which is the parent Hamiltonian for the GS wave functional Eq.~\eqref{eq:gsheight} and is given by
\begin{equation}
    H^{(m)}  = \gamma \int_0^L dx \; \mathcal{Q}^\dagger_m(x) \mathcal{Q}_m(x) \, ,
\label{eq:hamilhm}
\end{equation}
where $\gamma$ is an overall dimensionful constant.
The creation and annihilation operators $\mQ_m^\dagger(x)$ and $\mQ_m(x)$ are defined as
\begin{align}
\mathcal{Q}_m^\dagger(x)  =& \, \frac{1}{\sqrt{2}}\left(-\frac{\delta}{\delta \phim}  + (-1)^{m+1}\kappa\ \partial_x^{2(m+1)} \phim  \right) \nn \\
\mathcal{Q}_m(x)  =& \, \frac{1}{\sqrt{2}}\left(\frac{\delta}{\delta \phim} + (-1)^{m+1}\kappa\ \partial_x^{2(m+1)} \phim \right) \, ,
\label{eq:createannihildefn}
\end{align}
and satisfy the commutation relations
\beq
\label{eq:createannihilate}
[\mathcal{Q}_m(x), \mathcal{Q}_m^\dagger(y)] = (-1)^{m+1} \kappa\ \partial_x^{2(m+1)} \delta(x-y) \, .
\eeq
We can directly verify that the wave functional $\Phi^{(m)}_0 [\phim]$ Eq.~(\ref{eq:gsheight}) is a ``frustration-free" ground state of the Hamiltonian $H^{(m)}$ Eq.~(\ref{eq:hamilhm}) by noting that (see Eqs.~(\ref{eq:notfreeparticle}) and (\ref{eq:Ederivative}))
\begin{equation}
    \frac{\delta}{\delta \phim}\Phi^{(m)}_0[\phim] = -(-1)^{m+1} \kappa \left( \partial_x^{2(m+1)} \phim \right) \Phi^{(m)}_0[\phim],
\end{equation}
resulting in 
\begin{equation}
    \mathcal{Q}_m(x) \Phi^{(m)}_0 [\phim ]=0\;\;\forall x.
\end{equation}
Up to a constant (infinite) energy shift, the continuum Hamiltonian Eq.~(\ref{eq:hamilhm}) can be brought to more standard form~\cite{ardonne2004, Chen_2017a}
\begin{equation}
H^{(m)} = \gamma \int_0^{L} dx \left[
\frac{1}{2} \left( \Pi^{(m)} \right)^2
+
\frac{\kappa^2}{2}
\left(\partial_x^{2(m+1)} \phim\right)^2 
\right] \, ,
\label{eq:continuumhamil}
\end{equation}
where  $\Pi^{(m)}(x) = i\delta / \delta \phim(x)$ is the canonical momentum which satisfies $[\phim(x) , \Pi^{(m)}(y)] = i \delta(x-y)$. 

We observe that the Hamiltonian Eq.~(\ref{eq:continuumhamil}) is invariant under a polynomial shift of the form
\begin{equation}
    \phim(x) \rightarrow \phim(x) + P^{(2m+1)}(x),
\label{eq:polyshifthamil}
\end{equation}
where $P^{(2m + 1)}(x)$ is a polynomial in $x$ of degree $\leq (2m + 1)$.
That is, it has additional symmetries beyond just the $m^{th}$ multipole moment, as is typical of continuum RK-Hamiltonians~\cite{fradkin2004RK}.
However, using Eq.~(\ref{eq:nthmoment}), it is straightforward to see that the transformation Eq.~(\ref{eq:polyshifthamil}) changes the quantum number sector of the system.  This shows that the continuum Hamiltonian Eq.~(\ref{eq:continuumhamil}) is the same across all quantum number sectors, further implying that the ground state sector is extensively degenerate.
Now that we have established the form of the continuum Hamiltonian, we can study its lowest energy excited states to derive the dispersion relation and the gap. 
Using Eqs.~(\ref{eq:hamilhm}) and (\ref{eq:createannihilate}), the excited states $\Phi_k[\phim]$ can be written as
\begin{equation}
\Phi_{k}[\phim(x)]= \int_0^L dx \, f^{(m)}(k x) \, \mathcal{Q}^\dagger_m (x) \Phi_0[\phim (x)],
\end{equation}
where the mode function $f^{(m)}(kx)$ is determined by the boundary constraints on the height field $\phim(x)$, where $k$ is the momentum of the mode. For large system sizes, we expect that deep within the bulk $f^{(m)}(kx) \sim e^{ikx}$~\cite{Chen_2017b} from which we obtain the dispersion relation
\begin{equation}\label{eq:disp}
H^{(m)} \Phi_k[\phim ]= \gamma \kappa k^{2(m+1)} \Phi_k[\phim ] \;.
\end{equation} 
For a finite system of size $L$, we thus expect the gap $\Delta^{(m)}$ of $H^{(m)}$ to scale as
\begin{equation}
    \Delta^{(m)} \sim  \frac{1}{L^{2(m+1)}}.
\label{eq:gap}
\end{equation}

For charge-conserving systems ($m=0$), we see that the continuum height field approach correctly reproduces the scaling of the spin-wave dispersion relation of the Heisenberg model discussed in Sec.~\ref{sec:multipole}. 
More generally, we can further lower-bound the scaling of the Thouless time $\tth^{(m)}$ for a system of size $L$ conserving the $m^{th}$ multipole moment as follows:
\beq
\label{eq:tth}
\tth^{(m)} \gtrsim L^{2(m+1)} \;.
\eeq
Due to the polynomial shift symmetry (Eq.~(\ref{eq:polyshifthamil})) of the continuum Hamiltonian, we expect that this scaling of the Thouless time is independent of the quantum number sector. Eq.~\eqref{eq:tth} is one of the main results of this paper as it encodes the subdiffusive scaling of the Thouless time in systems with higher moment conservation laws. These results, obtained analytically through the generalized height representation developed herein, are validated by the numerical analysis performed on dipole conserving FRQCs (see Sec.~\ref{sec:multipole}).
The applicability of our continuum analysis of $\hrk$ extends beyond the context of random quantum circuits and is directly pertinent to the study of classical cellular automata with conserved higher moments. Such automata were studied in Refs.~\cite{feldmeier2020, morningstar2020} and are equivalent to the circuit $\hM$.   
To further test the validity of the continuum Hamiltonian obtained in Eq.~\eqref{eq:continuumhamil}, we can compute the two-point spin correlations using Eq.~(\ref{eq:generalizedheight}):
\beq
\langle \rho(x,t) \rho(0,0)\rangle_H  \propto  \frac{1}{\left(\kappa t \right)^{1/2(m+1)} } \mathcal{F}\left(\frac{x^{2(m+1)}}{\kappa t} \right) \;,
\label{eq:correlation}
\eeq
where $\rho(x,t) = e^{-iH^{(m)} t} \rho(x) e^{iH^{(m)}t}$, $\langle \cdot \rangle_H = \int \mathcal{D}\phim  (\cdot) \exp(-i\int dx\, dt\, H^{(m)}[\phim ])$, and $\mathcal{F}$ is a hypergeometric scaling function.
Eq.~(\ref{eq:correlation}) is in agreement with scalings obtained from numerical calculations and hydrodynamic considerations in Ref.~\cite{feldmeier2020}.
%
%
\section{Higher dimensional circuits}
\label{sec:higherdims}
We now briefly discuss extensions of our results to constrained FRQCs in dimensions $d > 1$, and in particular, systems on a hypercubic lattice that conserve all components of the $m^{th}$ multipole moment. 
First, we note that the discussions in Secs.~\ref{sec:FRQC} and \ref{sec:mapping} generalize directly \textit{mutatis mutandis} to higher dimensions.
We start with a $d$-dimensional spatially-random FRQC $\hW$ acting on a set of sites carrying color and spin DOFs, with the local Hilbert space given by $\mathcal{H}_{\textrm{loc}} = \mathbb{C}^q \otimes \mathbb{C}^{2s+1}$.
The circuit $\hW$ takes the form of Eq.~(\ref{eq:wdef}), comprising several layers $\{\hW_a\}$ composed of local unitary gates $\hU_{[\cdot]}$. 
The layers of $\{\hW_a\}$ are arranged in a ``Trotterized'' form: (i) For a given  $\{\hW_a\}$, $\hU_{[\cdot]}$  commute with each other, and all sites are being acted upon by exactly one gate; and (ii) each group of neighboring sites will be acted by a $\hU_{[\cdot]}$ in some $\{\hW_a\}$ in $\hW$ once and only once.
An example of a two-dimensional system with charge conservation will be provided below in Eq.~\eqref{eq:w_2dim_charge}.
%

%
%
%
%
%
%
%

\begin{figure}
    \centering
    \includegraphics[width=0.9\columnwidth]{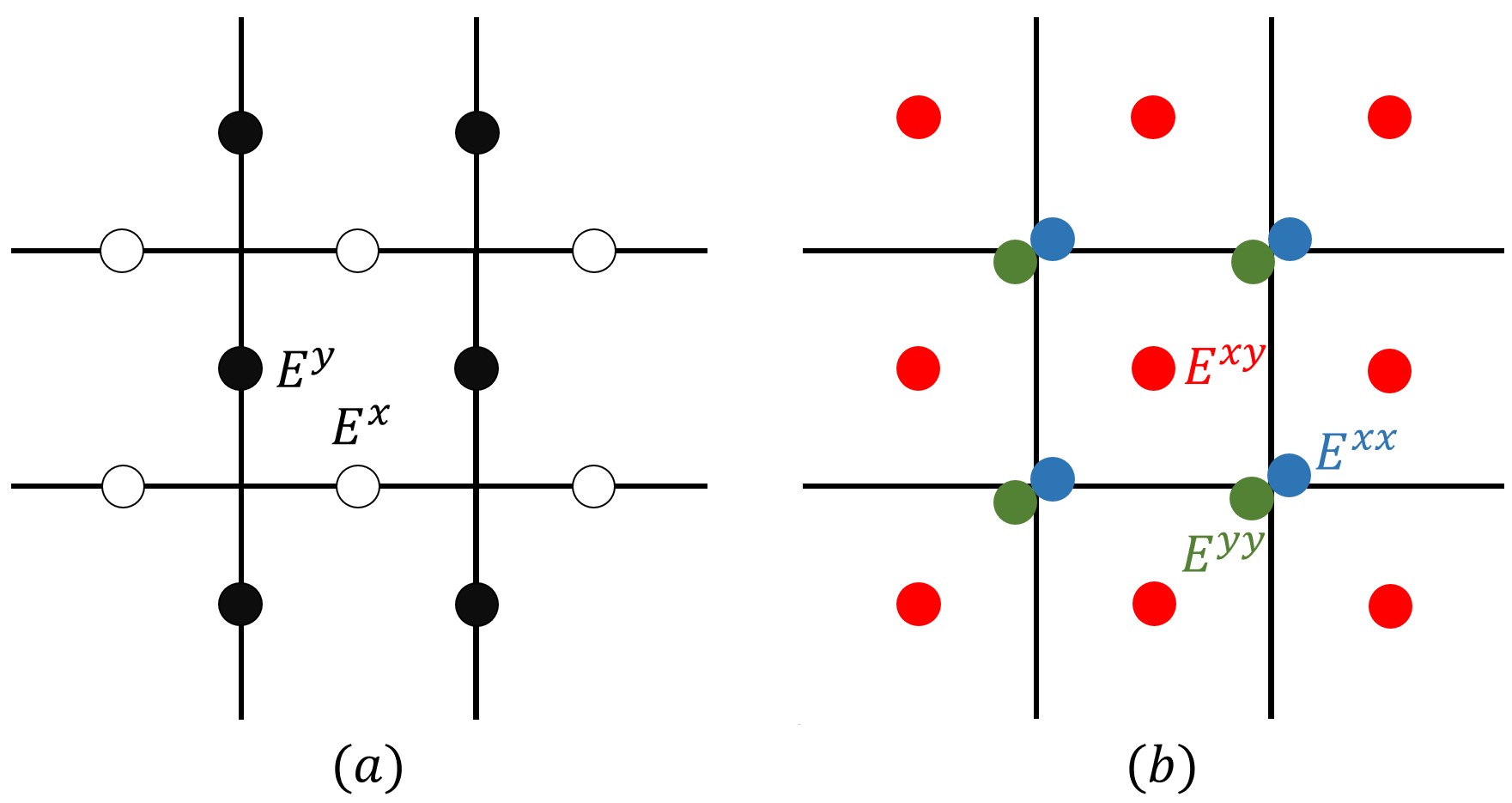}
    \caption{(a) Charge conserving case: independent components $E^x$ and $E^y$ live on links of the square lattice. (b) Dipole conserving case: diagonal components $E^{xx}, E^{yy}$ of the generalized electric field tensor live on each site while the off-diagonal component $E^{xy} = E^{yx}$ lives on each plaquette.}
    \label{fig:tensors}
\end{figure}

%
As before, we impose symmetries or local constraints on the spin DOFs and take the large-$q$ limit in the color DOFs; thus, the local gates $\hU_{[\cdot]}$ have the block-diagonal forms shown in Eq.~(\ref{eq:udef}).  
Using techniques directly generalized from Appendix~\ref{app:RKmapping}, we find that in the $q \rightarrow \infty$ limit, the SFF is expressed as Eq.~(\ref{eq:Kexpand}), where $\hM$ is a bistochastic matrix that retains the geometry of the original circuit $\hW$ but with its unitary gates $\hU_{[\cdot]}$ replaced by bistochastic matrices $\hm_{[\cdot]}$ of the form of Eq.~(\ref{eq:m_gates}), with the same transitions between local spin configurations as the original circuit. 
Following the arguments in Sec.~\ref{sec:mapping}, the Thouless time of the FRQC is related to the second largest eigenvalue of $\hM$ $\exp\left(-\Delta^{(\mk)}_{\hM}\right)$ within a given quantum number sector or Krylov subspace $\mk$ according to Eq.~(\ref{eq:tthbound}).  
Further, as discussed in Sec.~\ref{sec:mapping}, we can approximate the second largest eigenvalue of $\hM$ by the gap of an emergent RK-Hamiltonian (see Eq.~(\ref{eq:gapapprox})) that is a sum of local terms $\Pi_{[\cdot]}$ obtained from $\hm_{[\cdot]}$, following Eq.~(\ref{eq:HRKdefn}). 
This gap can then be used to deduce the scaling of the Thouless time $\tth$ with the system size. 
In what follows, we will be interested in $d$-dimensional systems that conserve all components of the $m^{th}$ multipole moment. We also restrict ourselves to hypercubic lattices with OBC in all directions, with coordinates labelled by a $d$-dimensional vector $\vx = (x_1, \cdots, x_d)$.  
The $m^{th}$ multipole moment operators are given by rank-$(m+1)$ symmetric tensors $\hQ^{i_1 \cdots i_m}_m$, defined as
\begin{equation}
    \hQ^{i_1 \cdots i_m}_m = \sumal{\vx}{}{x_{i_1} \cdots x_{i_m} S^z_{\vx}} \, ,
\label{eq:multipolehigherd}
\end{equation}
where $S^z_{\vx}$ is the Pauli-$Z$ matrix acting on site $\vx$, the indices of the tensor $\{i_j\}$ ($1 \leq i_j \leq d$) represent the $d$ lattice directions, and the summation runs over all sites of the hypercubic lattice. Note that when $d = 1$, we recover Eq.~(\ref{eq:multipole}). Quantum numbers associated with the operators $\{\hQ^{i_1 \cdots i_m}_m\}$ will be denoted by $\{Q^{i_1 \cdots i_m}_m\}$.
For example, the expressions for charge ($m = 0$) and dipole moment ($m = 1$) are
\begin{equation}
    \hQ_0 = \sumal{\vx}{}{S^z_{\vx}},\;\; \hQ^i_1 = \sumal{\vx}{}{x_i S^z_{\vx}} \, .
\label{eq:chargedipolehigherd}
\end{equation}
We now illustrate the above with an example and calculate $\tth$ for an FRQC composed of charge-conserving gates acting on spin-1/2 DOFs living on neighboring sites of a square lattice (with OBC). 
The circuit $\hW$ in this case can be implemented in four layers:
\begin{eqnarray}\label{eq:w_2dim_charge}
    \hW = &\bigotimesal{x\ \textrm{odd},\ y}{}{\hU_{[(x, y), (x + 1, y)]}} \bigotimesal{x\ \textrm{even},\ y}{}{\hU_{[(x, y), (x + 1, y)]}} \nn \\
    &\bigotimesal{y\ \textrm{odd},\ x}{}{\hU_{[(x, y), (x, y + 1)]}}\bigotimesal{y\ \textrm{even},\ x}{}{\hU_{[(x, y), (x, y + 1)]}} \, ,
\end{eqnarray}
where $\hU_{[(x, y), (x + \alpha, y + \beta)]}$ denotes the local charge-conserving gate acting on the rectangular region bounded on the bottom left and top right by the vertices $(x, y)$ and $(x + \alpha, y + \beta)$ respectively.
Similarly, the matrix $\hM$ has the structure
\begin{eqnarray}
    \hM = &\bigotimesal{x\ \textrm{odd},\ y}{}{\hm_{[(x, y), (x + 1, y)]}}\bigotimesal{x\ \textrm{even},\ y}{}{\hm_{[(x, y), (x + 1, y)]}}\nn \\
    &\bigotimesal{y\ \textrm{odd},\ x}{}{\hm_{[(x, y), (x, y + 1)]}} \bigotimesal{y\ \textrm{odd},\ x}{}{\hm_{[(x, y), (x, y + 1)]}},
\label{eq:Mform2d}
\end{eqnarray}
where each of the $\hm_{[(x, y), (x + \alpha, y + \beta)]}$ is a $4 \times 4$ matrix that has the form shown in Eq.~(\ref{eq:chargecons}).
Following Eqs.~(\ref{eq:HRKdefn}) and (\ref{eq:heisenberg}), the corresponding RK-Hamiltonian is the spin-1/2 ferromagnetic Heisenberg Hamiltonian in two dimensions:
\begin{equation}
    H_{\textrm{RK}} = \frac{1}{4}\sumal{\langle i, j\rangle}{}{\left(1 - \vec{\sigma}_i\cdot \vec{\sigma}_j\right)},  
\label{eq:HRK2d}
\end{equation}
where $\langle i, j\rangle$ represents nearest-neighboring sites on the square lattice.
Similar to the one-dimensional case, the Hamiltonian Eq.~(\ref{eq:HRK2d}) has a ferromagnetic ground state and its lowest energy excitations can be solved exactly; these are known to be spin-waves with a dispersion relation $\epsilon(k_x, k_y) = 2 \sin^2\left(k_x/2\right) + 2 \sin^2\left(k_y/2\right)$, where $k_x$ and $k_y$ represent the momenta of the spin wave in the $x$ and $y$ directions respectively. 
Furthermore, the Hamiltonian Eq.~(\ref{eq:HRK2d}) is $SU(2)$ symmetric so that the low-energy spectrum is the same within any of the $Q_0$ sectors. The gap within any $S_z$ sector thus scales as (if $L_x > L_y$)
\begin{equation}
    \Delta_{\textrm{RK}}(L_x, L_y) = \epsilon\left(k_x = \frac{2\pi}{L_x}, k_y = 0\right) \sim \frac{1}{L_x^2}.
\end{equation}
Following Eq.~(\ref{eq:gapapprox}), the Thouless time for a charge conserving system hence scales with the square of the longest linear-size of the system, consistent with expected results from diffusion. 
This discussion generalizes directly to charge conserving FRQCs acting on $d$-dimensional hypercubic lattices, where the emergent RK-Hamiltonian is the ferromagnetic Heisenberg Hamiltonian in $d$-dimensions with spin-wave excitations and the Thouless time scales as the square of the linear-size of the system. 
For dipole and higher multipole moment conserving systems, in general, or for charge-conserving FRQCs with higher spins or longer-range gates, the emergent RK-Hamiltonian is generically non-integrable.
Similar to the one-dimensional case, we hence consider systems with weak fragmentation~\cite{khemani2020}, take the continuum limit and resort to field theoretic arguments to obtain the gap scaling of the resulting Hamiltonians. 
Recall that the ground state of an RK-Hamiltonian is an equal superposition of all configurations within a given quantum number sector, similar to the one-dimensional case (see Sec.~\ref{sec:GS} and App.~\ref{app:continuumwf}); in the continuum limit, the ground state wavefunctional in $d$-dimensions is then
\begin{equation}
    \Phi^{(m)}_0 [\rho(\vx)] = \frac{1}{\sqrt{\mathcal{Z}}} 
\, e^{-\frac{\kappa}{2}\int{d^d\vx\  (\rho(\vx))^2}} \; \times \;\mathcal{G}[\rho(\vx)]
\label{eq:GSwfhigherd}
\end{equation}
where $\rho(\vx)$ is the charge density, $\mathcal{Z}$ is a normalization factor, and $\mathcal{G}[\rho(\vx)]$ enforces the global symmetry constraints i.e., it fixes the quantum numbers of the sector of interest. 
For example, for a dipole conserving system in $d$-dimensions with quantum numbers $\{Q_0, \{Q^j_1\}\}$, we have 
\begin{equation}
    \mathcal{G}[\rho(\vx)] = \delta\left(\int{\rho(\vx)} - Q_0\right) \prodal{j = 1}{d}{\delta\left(\int{x^j \rho(\vx)} - Q^j_1\right)} \, .
\label{eq:qnumhigherd}
\end{equation}
We now need to derive a continuum parent RK-Hamiltonian for the wavefunctional Eq.~(\ref{eq:GSwfhigherd}). 
To circumvent the global constraints in Eq.~(\ref{eq:GSwfhigherd}), we need some analog of the generalized height fields that we had introduced for one-dimensional systems in Sec.~\ref{sec:genheight}.
As emphasized in that section, the key role of the generalized height fields is to translate the global constraints in the wavefunctional into boundary constraints. 
As shown in Eq.~(\ref{eq:nthmoment}), for $m^{th}$ multipole conserving 1D systems in the continuum, this was accomplished by demanding that the generalized height fields $\phim(x)$ satisfy the generalized Gauss law Eq.~(\ref{eq:generalizedheight}). 
The natural analog of the $m^{th}$ generalized height fields $\phim(x)$ in higher dimensions are given by symmetric rank-$(m+1)$ tensor fields $\{\Em{j}(\vx)\}$, versions of which have previously been studied in the context of fracton models~\cite{polyshift2015, pretko2018fracton, gromov2019multipole, pretko2020review}.
To recast the global symmetry constraints enforcing the conservation of the $n^{th}$ multipole moments ($n<m$) in terms of boundary constraints on the tensor fields, we impose the following generalized Gauss law on the rank-$(m+1)$ tensor fields:
\begin{equation}
    \partial_{j_0}\cdots \partial_{j_m} \Em{j} (\vx) = \rho(\vx),
\label{eq:generalizedgausssym}
\end{equation}
where we sum over repeated indices.
We can then express the conserved quantities $\{Q^{i_1 \cdots i_m}_m\}$ in terms of boundary constraints on the tensor fields. 
For example, in charge conserving systems ($m = 0$), Eq.~(\ref{eq:generalizedgausssym}) reduces to the usual Gauss law for electric fields $\partial_{j_0} E^{j_0}(\vx) = \rho(\vx)$, and the total charge $Q_0$ can be expressed as
\begin{equation}
    Q_0 = \int{d^d \vx\ \partial_{j_0} E^{j_0}} = \oint{d n_{j_0}\ E^{j_0}},
\label{eq:Q0charge}
\end{equation}
where $d n_{j_0}$ represents the ``area" element on the boundary of the system, and we have used integration by parts along with Stokes' theorem.
Similarly, in dipole conserving systems, Eq.~(\ref{eq:generalizedgausssym}) reduces to the generalized Gauss law for rank-2 symmetric tensor fields $\partial_{j_0} \partial_{j_1} E^{j_0 j_1}(\vx) = \rho(\vx)$ and the total charge $Q_0$ and dipole moments $\{Q^i_1\}$ can be expressed as~\cite{fractonreview2}
\begin{eqnarray}
    &Q_0 = \int{d^d \vx\ \partial_{j_0} \partial_{j_1} E^{j_0 j_1}} = \oint{d n_{j_0}\ \partial_{j_1} E^{j_0 j_1}} \, , \nn \\
    &Q^i_1 = \int{d^d \vx\ x^i \partial_{j_0} \partial_{j_1} E^{j_0 j_1}} = \oint{d n_{j_0}\ \left(x^i \partial_{j_1}E^{j_0 j_1} - E^{i j_0}\right)}.\nn \\
\label{eq:Q0Q1dipole}
\end{eqnarray}
It is straightforward to show that a general expression for the $n^{th}$ multipole moment can also be derived in terms of boundary integrals of rank-$(m+1)$ symmetric tensor fields $\{\Em{j}(\vx)\}$ for any $m \geq n$, although the general expressions are rather tedious to show here and are not particularly illuminating. 
%
%
%
%
%
Thus, for a system with $m^{th}$ multipole moment conservation in all directions, we work in terms of rank-$(m+1)$ symmetric tensor fields, with the ground state wavefunctional Eq.~(\ref{eq:GSwfhigherd}) re-expressed as
\begin{eqnarray}\label{eq:gshigherDmult}
\Phi^{(m)}_0 [\{E^{i_0 \cdots i_m}(\vx)\}] &= \frac{1}{\sqrt{\mathcal{Z}}} 
\, e^{-\frac{\kappa}{2}\int d^d \vx\  (\partial_{j_0} \cdots \partial_{j_m} \Em{j} )^2} \nn \\
&\times \;\mathcal{B}[\{\Em{j}(\vx)\}] \;, 
\label{eq:GSwfhigherdtensor}
\end{eqnarray}
where $\mathcal{B}[\{E^{i_0 \cdots i_m}(\vx)\}]$ represents a boundary constraint on the fields $\{E^{i_0 \cdots i_m}(\vx)\}$ that fixes the quantum number sectors corresponding to all the $n^{th}$ multipole moments for $n \leq m$.
We now proceed to derive the expression for the parent RK-Hamiltonian corresponding to the wavefunctional Eq.~(\ref{eq:gshigherDmult}). 
The derivation closely follows the one-dimensional case discussed in Sec.~\ref{sec:hamiltonian}. 
The crucial idea is that in the long-wavelength limit, the Markov process corresponding to the RK-Hamiltonian of an $m^{th}$ multipole conserving system is simply the independent Langevin dynamics of each component of the rank-$(m+1)$ tensor field at each point. 
We can intuitively understand this on a  two-dimensional square lattice, where we label the two directions by $\hat{x}$ and $\hat{y}$. 
The generalized Gauss law of Eq.~(\ref{eq:generalizedgausssym}) is then discretized appropriately, and acts locally around each site of the lattice. 
In charge conserving systems, the rank-$1$ electric field $E^i$ has two components $E^x$ and $E^y$, which can be thought of as DOFs on the links of the lattice along the $\hat{x}$- and $\hat{y}$-directions respectively (see Fig.~\ref{fig:tensors}a). 
As a consequence of the discrete Gauss law, any nearest-neighbor charge conserving process along a link in the $\hat{x}$ (resp. $\hat{y}$) direction only modifies the fields $E^x$ (resp. $E^y$) on that link, whereas the electric fields far away remain unchanged.
In the continuum, such processes are modeled by the independent Langevin dynamics of $E^x$ and $E^y$ on each link. 
Similarly, in dipole conserving systems on a lattice, the rank-2 symmetric tensor field has three independent components: $E^{xx}$, $E^{yy}$, $E^{xy} = E^{yx}$.
The components $E^{xx}$ and $E^{yy}$ are DOFs on the vertices of the square lattice whereas $E^{xy} = E^{yx}$ are DOFs living on plaquettes of the square lattice (see Fig.~\ref{fig:tensors}).
As a consequence of the discrete generalized Gauss law, various local dipole conserving processes that occur independently result in independent fluctuations of these tensor fields.   
Furthermore, after coarse graining, we expect that the fluctuations in each component of the local fields will be Gaussian and that the fluctuations of different components of the tensors $\Em{j}$ will be uncorrelated.

Using the expression Eq.~(\ref{eq:gshigherDmult}), (in Appendix \ref{app:continuumhamilhigherd}) we derive the following expression for the continuum Hamiltonian:
\begin{equation}
    H^{(m)} = \gamma \int{d^d\vx\ \left(\mathcal{Q}^\dagger_m(\vx)\right)^{l_0 \cdots l_m} (\mathcal{Q}_m(\vx))_{l_0 \cdots l_m}}, 
\label{eq:continuumhamilhigherd}
\end{equation}
where repeated indices are summed over, and $(\mQ^\dagger(\vx))^{l_0 \cdots l_m}$ and $(\mQ(\vx))_{l_0 \cdots l_m}$ are respectively creation and annihilation operators for the fluctuations of the component $\Em{l}$; their explicit expressions are given by Eq.~(\ref{eq:realspacealgebrahigherd}).  
Note that we obtain separate creation and annihilation operators for each component of the rank-$(m+1)$ tensor since their fluctuations are independent.
Further, using the properties of these operators shown in Eq.~(\ref{eq:realspacealgebrahigherd}), the lowest excited state $\Phi^{(m)}_{\vk}(\Em{j}(\vx))$ with momentum $\vk$ is given by (see Eq.~(\ref{eq:excitedwfrealspacehigherd}))
\begin{equation}
    \Phi^{(m)}_{\vk} = \int{d^d\vx\ e^{i \vk\cdot \vx} (k_{l_0} \cdots k_{l_m})(\mQ^{\dagger} _m(\vx))^{l_0 \cdots l_m} \Phi^{(m)}_0},
\end{equation}
where repeated indices are summed over, and we have supressed the arguments in $\Phi^{(m)}_{\vk}$ and $\Phi^{(m)}_0$.
$\Phi^{(m)}_{\vk}$ can also be shown to satisfy 
\begin{equation}
    H^{(m)} \Phi^{(m)}_{\vk} = \gamma \kappa\ (\sumal{l = 1}{d}{k_l^2})^{m+1} \Phi^{(m)}_{\vk}. 
\end{equation}
For a system with linear-size $L_j$ in the $j^{th}$ direction, we thus expect the gap to scale as (assuming $L = \max_j(L_j)$)
\begin{equation}
    \Delta^{(m)} \sim \frac{1}{L^{2{(m+1)}}}, 
\end{equation}
thereby showing that the Thouless time follows the scaling of Eq.~(\ref{eq:tth}) i.e., the Thouless time for multipole conserving circuits in higher dimensions follows the same subdiffusive scaling with the linear extent of the system as that of one-dimensional multipole conserving circuits.
While we have primarily focused on systems that conserve all components of the $m^{th}$ multipole moment, this formalism directly generalizes to systems where only a few components of $m^{th}$ multipole moments are conserved.  
Such a setting is directly relevant to many physical systems, for instance in recent experiments that impose dipole moment conservation only along a single direction by subjecting the system to a strong electric field in that particular direction.
Continuum wavefunctions of the form Eq.~(\ref{eq:GSwfhigherdtensor}) for such systems can also be expressed in terms of tensor fields that obey anisotropic versions of the Gauss law Eq.~(\ref{eq:generalizedgausssym})~\cite{gromov2019multipole}.
For example, in a two-dimensional system with charge conservation in the $x$-direction and dipole moment conservation in the $y$-direction, we obtain
\begin{equation}
    \partial_x E^x + \partial_y \left(\partial_x E^{yx} + \partial_y E^{yy}\right) = \rho(\vx). 
\label{eq:anisotropicgauss}
\end{equation}
Following similar ideas as in the isotropic case, it is then straightforward to derive expressions for the continuum Hamiltonian similarly to Eq.~(\ref{eq:continuumhamilhigherd}), which corresponds to Langevin dynamics of each of the tensor components involved, and to then derive the scaling of the Thouless time.
We find that the Thouless time for the entire system is dominated by the highest multipole moment conserved, i.e. $\tth \sim L^{2(m+1)}$ if some component of the $m^{th}$ multipole moment (but none higher) is conserved, consistent with intuition and experimental observations~\cite{bakr2019}.
%
%
%
\section{Concluding Remarks}
\label{sec:cncls}
In this paper, we have studied the spectral statistics, as encoded in the SFF $K(t)$, for spatially-extended constrained many-body quantum chaotic systems, focusing on FRQCs with conserved higher moments, such as the dipole moment. As one of the key results of this paper, we have established a series of relations between $K(t)$ in the $q\to\infty$ limit, a \textit{classical} stochastic circuit $\hM$, and an emergent RK-Hamiltonian, such that the inverse gap of this RK-Hamiltonian lower bounds the Thouless time $\tth$ of the underlying FRQC. As we have shown here, the relation between $\tth$ and $\Delta_{\mathrm{RK}}$ proves particularly efficacious, since it relates a dynamical property of the FRQC to the low-energy physics of a sign-problem-free quantum Hamiltonian.

We emphasise that these relations are valid for \textit{generic} local FRQCs with on-site Abelian symmetries or dynamical constraints, not only those with conserved higher moments of charge. For example, we can consider circuit implementations of other fragmented models~\cite{yang2019} or study an FRQC inspired by the Rydberg blockade~\cite{lesanovsky2011many, bernien2017probing}, also known as the PXP model~\cite{turner2018}.
The latter is implemented by taking e.g., $\ell=3$ site local gates with the only non-trivial dynamics contained within a $2\times 2$ block connecting the $\ket{\downarrow \downarrow \downarrow}$ and $\ket{\downarrow \uparrow \downarrow}$ states.
The resulting Floquet operator $\hW$ has no conserved quantities besides the (quasi)-energy, but fragments into dynamically disconnected subspaces; the largest of these corresponds to the constrained Hilbert space most often discussed in the context of quantum many-body scar dynamics~\cite{turner2018}.

We have verified these general results on circuits with higher conserved moments, which generically exhibit Hilbert space fragmentation. Working in the $q\to\infty$ limit, we derived the corresponding stochastic circuit $\hM$ and emergent RK-Hamiltonian $\hrk$ for both charge and (weakly fragmented) dipole conserving systems. Our numerical study of these systems suggests a universality in the scaling of $\tth$ with system size, specifically, we predict diffusive scaling $\tth \sim L^2$ for charge conserving systems and subdiffusive behavior $\sim L^4$ for dipole conserving systems, regardless of the microscopic details of the underlying circuit.
Further evidence for this scaling is given by continuum field theoretic descriptions of the emergent RK-Hamiltonians for multipole conserving FRQCs in terms of generalized height fields. By analytically computing the dispersion relation for the resultant field theories, we find that $\tth \sim L^{2(m+1)}$ in circuits that conserve the $m^{th}$ multipole moment, consistent with numerical results for charge and dipole conserving systems. 
We further generalize our formalism to higher dimensions, where we derive continuum field theories for emergent RK-Hamiltonians for systems that conserve dipole and higher multipole moments. We obtain the same scaling of the Thouless time with the largest linear size of the system i.e., $\tth \sim L^{2(m+1)}$ for circuits that conserve any component of the $m^{th}$ multipole moment (but none higher) in any number of dimensions, consistent with expectations from the one-dimensional result.

Our work opens many exciting avenues for future research: here, we have only focused on the class of multipole conserving circuits which exhibit weak fragmentation. Dynamics in strongly fragmented systems, where typical initial states are ETH-violating, is highly constrained; nevertheless, such systems exhibit large Krylov subspaces which eventually thermalize~\cite{moudgalya2019krylov}. The scaling of $\tth$ within such subspaces remains to be understood and may lead to distinct continuum field theories than those we have introduced for weakly fragmented systems.
Another interesting avenue to explore is extending our formalism to incorporate non-Abelian symmetries, for which the nature of transport and thermalization is currently being debated~\cite{protopopov2020nonabelian,yang2020anomalous, glorioso2020hydrodynamics}.
We note that the large-$q$ diagrammatics, and therefore the mapping to a classical bistochastic circuit and RK-Hamiltonian, have so far only been developed for the two-point SFF $K(t)$. Other observables, such as the second Renyi entropy and out-of-time-order correlator, can also be mapped to stochastic classical dynamics upon ensemble averaging, and will be discussed in forthcoming work. Pushing these ideas further presents an important but technically-demanding theoretical challenge. More straightforward is extending our results to circuit geometries besides the brick-wall structure considered here as well as to other RMT symmetry classes. 

More pressing, however, is building a systematic understanding of FRQCs at finite-$q$, to delineate those features which are an artefact of the $q\to\infty$ limit from those which are more generic properties of constrained random circuits. Numerically investigating finite-$q$ circuits remains prohibitive, particularly in the context of higher moment conserving circuits which already require large ($\ell \geq 4$) local gates. Analytically, one could attempt to keep track of diagrams at next to leading order in the large-$q$ expansion to better quantify deviations of the SFF from the strict $q\to\infty$ limit. We leave the development of such analytical techniques to future work.


\begin{acknowledgements}
We are particularly grateful to Shivaji Sondhi for enlightening discussions. We also acknowledge useful conversations with Nathan Benjamin, John Chalker, Andrea De Luca, Alan Morningstar, and Pablo Sala. A. P. was supported in part with funding from the Defense Advanced Research Projects Agency (DARPA) via the DRINQS program. The views, opinions and/or findings expressed are those of the authors and should not be interpreted as representing the official views or policies of the Department of Defense or the U.S. Government. A. C. is supported in part by the Croucher foundation. A. P. and A. C. are supported by fellowships at the PCTS at Princeton University.  D. A. H. is supported in part by DOE grant DE-SC0016244. 

\end{acknowledgements}


\bibliography{library}

\clearpage
\appendix
\onecolumngrid


\section{Mapping $K(t)$ to a classical Markov circuit}
\label{app:RKmapping}

\begin{figure}[b]
	\includegraphics[width = 0.85\columnwidth]{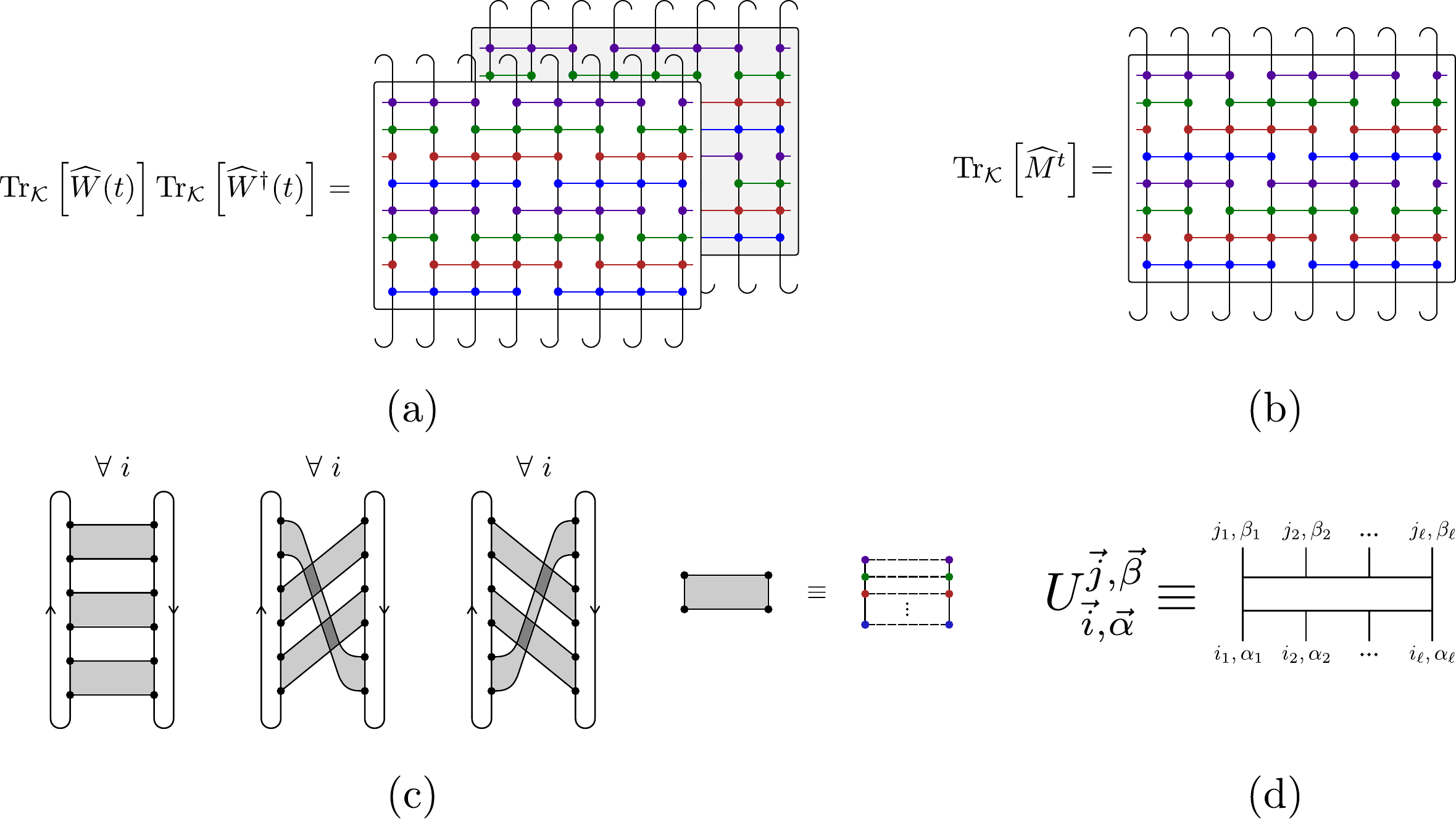} 
	\caption{
(a) A schematic illustration of the diagrammatic representation of $K(t;\mathcal{K})$ (adapted from Ref.~\cite{cdc1}) for a FRQC with $\ell = 4$ (which is appropriate for the minimal spin-1 dipole-conserving FRQC with weak fragmentation). The white and grey sheets represent $\mathrm{Tr}_{\mathcal{K}}[\hW(t)]$ and $\mathrm{Tr}_{\mathcal{K}}[\hW^\dagger(t)]$ respectively. Space runs horizontally and time runs vertically. Curly lines on the top and the bottom of the sheet represents traces over the dof at each site. Gates with different support are denoted with different colors. 
(b) The diagrammatic representation of $\mathrm{Tr}_{\mathcal{K}} [\hM^t]$ where, unlike $\hW$, $\hM$ is a non-random circuit that acts only on the spin degrees of freedom of the FRQC $\hW$. Each $\ell$-gate in $\hM$ is a block-diagonal bistochastic matrix defined in Eq.~\eqref{eq:m_gates}. All charges are preserved after the action of every $\ell$-site gates in $\hM$.
(c) The three leading diagrams in the diagrammatic expansion of $K(t;\ms)$  at $t=3$ in the limit of large-$q$~\cite{cdc1}.
Note that for each diagram, every site $i$ takes the \textit{same} configuration.
The grey ribbons correspond to a ladder of $\ell$ number of contractions between unitaries and their conjugates, represented by colored dots. (d) The diagrammatic representation of $U_{\vec{i} ,\vec{\alpha}}^{\vec{j},\vec{\beta}}$, where the Roman (Greek) indices correspond to color (spin) degrees of freedom. \label{fig:app_a} }
\end{figure}

In this Appendix, we generalize the calculation of SFF $K(t; \mathcal{K})$ performed in Ref.~\cite{friedman2019} for an FRQC with conserved U(1) charge to FRQCs with arbitrary symmetries or local constraints, such as dipole moment conservation.
Specifically, we show that in the $q \rightarrow \infty$ limit, $K(t; \mathcal{K})$ is mapped to the trace of the $t$-th power of a bistochastic matrix.
The ensemble average in $K(t; \mathcal{K})$, defined in Eq.~\eqref{eq:K_krylov} and illustrated in Fig.~\ref{fig:app_a}(a), can be evaluated as a sum of diagrams, each of which corresponds to a possible pairing between unitaries and their complex conjugates.
In the limit of large-$q$, diagrams with the same ``cyclical'' pairing at all sites dominate the sum (see illustration in Fig.~\ref{fig:app_a}(c) and  Ref.~\cite{cdc1} for details).
There are $t$ such diagrams, which we call ``Gaussian'' since they can be evaluated using the Wick contractions between unitaries and their conjugates. For $\ell$-site unitaries $U$ and their conjugates $U^\ast$, shown in Fig.~\ref{fig:app_a}(d), we have
\begin{align}
\label{eq:wick}
\left\langle U_{\vec{i} ,\vec{\alpha}}^{\vec{j},\vec{\beta}}
	U_{\vec{i}' ,\vec{\alpha}'}^{ \ast \vec{j}',\vec{\beta}'}
	\right\rangle
	&= \frac{1}{ d_{\vec{\alpha}} q^\ell}  \delta_{\vec{i}, \vec{i}'} \delta_{\vec{j}, \vec{j}'}  \delta_{\vec{\alpha}, 
	\vec{\alpha}'} \delta_{\vec{\beta}, \vec{\beta}'} \, ,   
 \quad  \quad \delta_{\vec{i}, \vec{i}'} = \prod_{a=1}^{\ell} \delta_{i_a, i'_a} \, ,
\end{align} 
where $\vec{i}  = (i_1, i_2, \dots, i_\ell)$ and $\vec{\alpha}  = (\alpha_1, \alpha_2, \dots, \alpha_\ell)$ are $\ell$-component vectors denoting respectively the color and spin degrees of freedom on their support.
Here, $d_{\vec{\alpha}}$ denotes the number of ``out-going" $\ell$-site spin configurations dynamically accessible to the ``in-coming" $\ell$-site spin configuration $\vec{\alpha}$.
That is, $d_{\vec{\alpha}} q^\ell$ is the size of the random matrix block in $\hW$ to which the basis state $(\vec{i},\vec{\alpha})$ belongs.

As an example, let us consider the simplest charge-conserving circuit~\cite{friedman2019}.
We have two-site gates ($\ell = 2$) and $\vec{\alpha} = (\alpha_1, \alpha_2)$, with $\alpha_i = \uparrow, \downarrow$.
The dynamically connected two-site configurations can be labelled by a conserved charge $Q_0 = \sum_{x=1}^{\ell}S^z_x$.
The configurations $(\uparrow, \uparrow )$ and $(\downarrow, \downarrow)$ have $Q_0(\vec{\alpha}) = +1$ and $Q_0(\vec{\alpha}) = -1$ respectively and are not dynamically connected to any other two-site configurations. Thus, these correspond to $d_{\vec{\alpha}}=1$.
However, the configurations $(\uparrow,\downarrow )$ and $(\downarrow, \uparrow)$ both have $Q_0( \alpha) = 0$, and are dynamically connected to each other. Thus we have $d_{\vec{\alpha}} = 2$.
Next, we can translate each of the $t$ diagrams into algebraic terms by using Eq.~\eqref{eq:wick} and summing over the color degrees of freedom.
Consider the diagram where the pairing between unitaries and their conjugates takes the form of the leftmost diagram in Fig.~\ref{fig:app_a}(c) at every site.
The sum over the colors precisely cancels out all factors of $q$ in Eq.~\eqref{eq:wick}~\cite{cdc1}. 
It then remains to sum over the spin degrees of freedom. 
Observe that the choice of spin DOFs in $\hW$ uniquely fixes those in $\hW^\dagger$ due to the Wick contractions Eq.~\eqref{eq:wick}.
Consequently, the sum over spins can be computed by finding all possible ways of assigning spins in the diagrammatic representation of $\Tr_{\mk}\left[\hW(t)\right]$ (Fig.~\ref{fig:app_a}(b), such that all charges $\{Q_a \}$ are preserved after the action of every $\ell$-site gate.
This sum over spins can be reproduced by $\textrm{Tr}_{\mk} \left[\hM^t\right] \,$ in Eq.~\eqref{eq:Kexpand}, where $\hM$ has the geometry of $\hW$ in Eq.~\eqref{eq:wdef} and contains non-random block-diagonal $\ell$-gates $\hm$, defined in Eq.~\eqref{eq:m_gates}. 
Note that $\hm$ contains $d_{\vec{\alpha}} \times d_{\vec{\alpha}}$ matrix blocks $\hm(d_{\vec{\alpha}})$ with entries $1/d_{\vec{\alpha}}$ to account for the factor $d_{\vec{\alpha}}$ in Eq.~\eqref{eq:wick}.
The overall factor of $|t|$ in Eq.~\eqref{eq:K_RK} arises from the following argument: any two of the $t$ contracted diagrams in Fig.~\ref{fig:app_a}(c) are related by a rotation of the arrowed loop on the right side of one of the diagrams i.e., the diagrams are topologically equivalent. As a result, the corresponding algebraic terms for all $t$ leading diagrams are identical. This concludes the derivation of Eq.~\eqref{eq:K_RK}.
%
%
%
%
\section{Continuum Limit of the RK Ground State wave function}
\label{app:continuumwf}
In this Appendix, we derive the continuum limit for the ground state wave function of multipole conserving RK-Hamiltonians $\hrk$, thereby allowing us to analyze these systems using field-theoretic techniques.
In the discrete setting, the ground state wave function Eq.~(\ref{eq:RKGS}) has the form
\begin{equation}
    \ket{\psi} = \sumal{\{s_n\} \in \Lambda}{}{\ket{\{s_n\}}} \, ,
\label{eq:GSwave function}
\end{equation}
where $\Lambda$ denotes the set of allowed spin configurations that are allowed within the quantum number sector or Krylov subspace of interest. 
As discussed in Sec.~\ref{sec:genheight}, the spins $\{s_n\}$ assume integer or half-integer values in $[-s, s]$. 
In the continuum limit i.e., when the lattice spacing $\Delta x \to 0$, under coarse-graining the distributions of $s_n$'s flows to Gaussian random variables with zero mean and variance $\sigma^2 \sim \dx$ as a consequence of the central limit theorem~\cite{majumdar2005brownian}. 
That is, the probability density $\mbP$ of a single spin $s_{n}$ is
\begin{equation}
    \mbP\left(s_{n}\right) \sim \exp\left(-\frac{s^2_{n}}{2\sigma^2}\right)\;\;\implies\;\;\; \langle s_{n} \rangle = 0,\;\;\; \langle s_{n} s_{n'}\rangle = \sigma^2 \delta_{n, n'},\;\;\; \sigma^2 = \frac{\dx}{\kappa} \, .
\label{eq:gaussiandist}
\end{equation}
Here, $\kappa$ effectively serves as the coarse-graining length scale and is determined by requiring that the continuum theory correctly reproduces long-distance correlation functions in the microscopic model, which can in principle be computed numerically. Next, by expressing the $s_n$'s in terms of the $m^{th}$ height fields and by using Eqs.~(\ref{eq:heightdefn}) and (\ref{eq:recursiveheight}), we obtain
\begin{equation}
    s_{n} = \phi^{(0)}_{n + \half} - \phi^{(0)}_{n - \half} = \phi^{(1)}_{n + 1} - 2\phi^{(1)}_{n} + \phi^{(1)}_{n - 1} = \cdots = \sumal{j = -\frac{m + 1}{2}}{\frac{m + 1}{2}}{(-1)^{j + \frac{m + 1}{2}} \binom{m + 1}{j + \frac{m + 1}{2}} \phim_{n + j}} \, ,
\label{eq:finitediff}
\end{equation}
while the joint probabilities satisfy
\begin{align}
    \mbP(\{s_{n}\}) \sim & \exp\left(-\frac{1}{2\sigma^2}\sumal{n}{}{s_{n}^2}\right) \times \mathcal{G}[\{s_n\}]\nn \\
    \implies\;\; \mbP(\{\phim_{n}\}) \sim & \exp\left(-\frac{1}{2\sigma^2}\sumal{n}{}{\left(\sumal{j = -\frac{m + 1}{2}}{\frac{m + 1}{2}}{(-1)^{j + \frac{m + 1}{2}} \binom{m + 1}{j + \frac{m + 1}{2}} \phim_{n + j}}\right)^2}\right) \times \mathcal{B}[\{\phim_n\}] \, ,
\label{eq:discreteconfigshigher}
\end{align}
where $\sigma^2 = \Delta x/\kappa$. $\mathcal{G}[\{s_n\}]$ and $\mathcal{B}[\{\phim_n\}]$ are shorthand for global and boundary constraints written in terms of spin and height variables respectively; these constraints are required to fix the quantum number sector, as discussed in Sec.~\ref{sec:genheight}. 
Note that the Jacobian of the transformation from $\{s_n\} \rightarrow \{\phim_{n}\}$ is $1$. 
In order to obtain a continuum limit in the case of $m^{th}$ height variables, we need to define height fields that scale with the lattice spacing as shown in Eq.~(\ref{eq:heightfields}). The probability density for a configuration of the $m^{th}$ height field in the continuum limit then reads
\begin{align}
    \mbP(\phim(x)) &\sim \lim_{\dx \rightarrow 0}{\exp\left(-\frac{\kappa}{2}\sumal{n}{}{\dx\ \left(\frac{\sumal{j = -\frac{m+1}{2}}{\frac{m+1}{2}}{(-1)^{j + \frac{m+1}{2}} \binom{m+1}{j + \frac{m+1}{2}} \phim_{n + j} (\dx)^{m}}}{(\dx)^{m+1}}\right)^2}\right)} \times \mathcal{B}[\{\phim_n\}] \nn \\
    &= \exp\left(-\frac{\kappa}{2}\int_0^L dx\ (\partial^{m+1}_x \phim(x))^2\right) \times \mathcal{B}[\{\phim(x)\}] \, ,
\label{eq:probhigher}
\end{align}
from which we obtain Eq.~\eqref{eq:gsheight} as the continuum limit of the ground state wave functional.
The full continuum wave function can be written as
\begin{equation}
    \ket{\Phi^{(m)}_0} = \frac{1}{\sqrt{\mZ}}\intal{\mathcal{B}[\phim(x)]}{}{\mD\phim\ \exp\left(-\frac{\kappa}{2}\int_0^L dx\ (\partial^{m+1}_x \phim(x))^2\right)\ket{\phim(x)}} \, .
\label{eq:wf}
\end{equation}
%
%
%
\section{Brief Interlude on SMF Decomposable Hamiltonians}
\label{app:smf}
In this Appendix, we discuss a generalization of the RK wave functions in Eq.~(\ref{eq:RKGS}) to states of the form 
\begin{equation}
    \ket{\psi_\textrm{SMF}} = \frac{1}{\sqrt{\mZ}}\sumal{\mC \in \Lambda}{}{p_{\mC} \ket{\mC}} \, ,
\label{eq:SMFwf}
\end{equation}
where the Hilbert space is spanned by basis states composed of an abstract set of configurations $\Lambda = \{\mC\}$, $p_{\mC}$ is the ``probability" associated with the configuration $\mC$, and $\sqrt{\mZ}$ is the normalization constant for the wave function. 
We recover the standard RK wave functions of the form Eq.~(\ref{eq:RKGS}) by setting $p_{\mC} = 1$ for all configurations $\mC$. 

As discussed in Refs.~\cite{henley2004, castelnovo2005}, wave functions of the form Eq.~\eqref{eq:SMFwf} are closely related to steady states of Markov processes satisfying detailed balance. Moreover, these generalized RK wave functions correspond to ground states of Hamiltonians which admit a ``Stochastic Matrix Form" decomposition and are of the following form:
\begin{eqnarray}
    &\hsmf = \sumal{\langle \mC, \mC' \rangle}{}{w_{\mC, \mC'} \hQ_{\mC, \mC'}}, \nn \\
    &\hQ_{\mC, \mC'} = \left[\frac{1}{p_{\mC}^2} \ket{\mC}\bra{\mC} + \frac{1}{p_{\mC'}^2} \ket{\mC'}\bra{\mC'} - \frac{1}{p_{\mC} p_{\mC'}}\left( \ket{\mC}\bra{\mC'} + \ket{\mC'}\bra{\mC}\right)\right]
    = \left(\frac{1}{p_{\mC}}\ket{\mC} - \frac{1}{p_{\mC'}}\ket{\mC'}\right)\left(\frac{1}{p_{\mC}}\bra{\mC} - \frac{1}{p_{\mC'}}\bra{\mC'}\right),
\label{eq:HSMF}
\end{eqnarray}
where $\langle \mC, \mC' \rangle$ represents a pair of configurations from the set $\Lambda$ that are connected under the action of the Hamiltonian.
In Eq.~(\ref{eq:HSMF}), $w_{\mC, \mC'}$'s are positive real numbers to ensure that $\hsmf \geq 0$.  
It is straightforward to verify by direct computation that $\ket{\psi_\textrm{SMF}}$ is a frustration-free ground state of $\hsmf$ since
\begin{equation}
    \hQ_{\mC, \mC'}\left(p_{\mC} \ket{\mC} + p_{\mC'} \ket{\mC'}\right) = 0 \;\;\; \implies \hsmf \ket{\psi_\textrm{SMF}} = 0. 
\label{eq:psiRKvanish}
\end{equation}
Typically, $p_{\mC}$ and $\mZ$ are represented as~\cite{castelnovo2005}
\begin{equation}
    p_{\mC} = e^{-\frac{\beta \mE_{\mC}}{2}},\;\;\;\mZ = \sumal{\mC \in \Lambda}{}{e^{-\beta \mE_{\mC}}}.
\label{eq:params}
\end{equation}
where $\mE_{\mC}$ is the ``energy" associated with the configurations, $\beta$ is the ``inverse-temperature" and hence the $\mZ$ resembles the partition function. We note that while the standard RK-Hamiltonian $\hrk$ Eq.~\eqref{eq:hrkstandardform} is proportional to the transition matrix of a Markov process, $\hsmf$ is related to a transition matrix $T$ of the Markov process through a similarity transformation, under which the ground state wave function remains of the form Eq.~\eqref{eq:SMFwf}.
%
%
\section{Continuum Limit of the RK Hamiltonian}
\label{app:continuumhamil}%
In this Appendix, we discuss the derivation of the continuum limit of the RK-Hamiltonian for multipole conserving systems.
We outline two distinct approaches, both of which lead to the same Hamiltonian.
\subsection{Regulator approach}
\label{app:continuumhamilreg}

Let us begin by observing that the continuum wave function Eq.~(\ref{eq:wf}) has the form Eq.~(\ref{eq:SMFwf}) i.e., it belongs to the class of ground states of SMF decomposable Hamiltonians. We thus anticipate that the continuum limit of the RK-Hamiltonian is SMF decomposable and can be brought to the form Eq.~\eqref{eq:HSMF}.

To make this correspondence precise, we identify the set of configurations $\Lambda$ with the set of height field configurations $\{\phim(x)\}$, while the inverse-temperature $\beta$, energy $\{\mE(\phim(x))\}$, and partition function $\mZ$ are respectively given by (see Eq.~(\ref{eq:params}))
\begin{equation}
    \beta = \kappa,\;\;\mE(\phim(x)) = \int_0^L dx\ \left(\partial^{m+1}_x \phim(x)\right)^2,\;\;\mZ = \intal{\mathcal{B}[\phim(x)]}{}{\mD\phim\ \exp\left(-\beta \mE(\phim_k)\right)} \, .
\label{eq:notfreeparticle}
\end{equation}
$\mathcal{B}[\phim(x)]$ is shorthand for the boundary constraints Eq.~(\ref{eq:nthmoment}) that fix the quantum number sector.
Note that in the case of systems involving $\ell$-site gates, two configurations of spins (height variables) $\mC = \{s_n\}$ ($\mC = \{\phim_{n}\}$) and $\mC' = \{s'_n\}$ ($\mC' = \{{\phim_{n}}'\}$) are connected under the action of the Hamiltonian only if the $\ell$ local spins ($\ell - 1$ height variables) satisfy the constraints imposed by the higher moment symmetries.
For example, in the case of a Hamiltonian with charge conservation,
\begin{eqnarray}
    &s_j + s_{j+1} = s'_{j} + s'_{j+1}\;\;\textrm{for some $j$}\;\;\textrm{and}\;\; s_n = s'_n\;\; \forall n,\;\; n \neq j, j+1 \nn \\
    &\iff \phi^{(0)}_{j + \half} = {\phi^{(0)}_{j + \half}}' + s_j - s'_j\;\;\textrm{for some $j$}\;\;\textrm{and}\;\; \phi^{(0)}_{n + \half} = {\phi^{(0)}_{n + \half}}'\;\; \forall n,\;\; n \neq j \, .
\label{eq:connectedcond}
\end{eqnarray}
We see that the Hamiltonian remains local when switching to the height representation, since the action of the Hamiltonian changes the height variable only at a single point.
While this is strictly true only for multipole conserving processes of the minimal gate-size, after coarse-graining we expect that the long-wavelength physics is captured by a Hamiltonian that only permits local fluctuations in the height field configurations. 
Consequently, after coarse-graining we can express the RK-Hamiltonian Eq.~\eqref{eq:hrkstandardform} for an $m^{th}$ multipole moment conserving system in the SMF-form Eq.~(\ref{eq:HSMF}):
\begin{equation}
    H^{(m)} = \sumal{\{\phim_n\}, \{{\phim_n}'\}}{}{\sumal{j}{}{\left(\prodal{n \neq j}{}{\delta_{\phim_n, {\phim_n}'}} \hQ_{\{\phim_n\}, \{{\phim_n}'\}} w_{\{\phim_n\}, \{{\phim_n}'\}}\right)}} \, .
\label{eq:RKlocal}
\end{equation}
Since in taking the continuum limit, we replace discrete spin variables by continuous Gaussian random variables with variance $\sigma^2 \sim \dx$, we replace sums in Eq.~(\ref{eq:RKlocal}) by integrals as follows
\begin{equation}
    \sumal{\{\phim_n\}, \{{\phim_n}'\}}{}{} \rightarrow \intal{}{}{\prodal{n}{}{d\phim_n d{\phim_n}'\ \mbP(\{\phim_n\}) \mbP(\{{\phim_n}'\})}} \, ,
\end{equation}
where $\mbP$ denotes the Gaussian probability density Eq.~(\ref{eq:discreteconfigshigher}). Discrete delta functions are replaced by continuous ones.
We then obtain the following expression for the continuum Hamiltonian $H^{(m)}$ \begin{equation}
    H^{(m)} = 
    \sumal{j}{}
    \int{\prodal{n \neq j}{}{d \phim_n d{\phim_n}'\ \delta(\phim_n - {\phim_n}'){\int{d\phim_j\ d{\phim_j}'\ \mbP(\{\phim_n\}) \mbP(\{{\phim_n}'\})}\ w_{\{\phim_n\}, \{{\phim_n}'\}} \hQ_{\{\phim_n\}, \{{\phim_n}'\}}}}}.
\label{eq:RKint}
\end{equation}
It is clear that with an appropriate choice of the regulator $w_{\{\phim_n\}, \{{\phim_n}'\}}$, we can write Eq.~(\ref{eq:RKint}) in the form
\begin{equation}
    H^{(m)} = \int{\prodal{n}{}{d \phim_n}\  \mbP(\{\phim_n\})^2\sumal{j}{}{\int{d\lambda_j  \exp\left(-\frac{\lambda^2_j}{\sigma^2}\right) \hQ_{\{\phim_n\}, \{{\phim_n}'\}}}}},\;\;\;\lambda_{j} \equiv \phi'_{j} - \phi_{j},
\label{eq:Hreg}
\end{equation}
which corresponds to the fluctuations of the height fields being Gaussian with variance $\sigma^2$. 
Note that since $\{\phim_n\}$ and $\{{\phim_n}'\}$ only differ in ${\phim_j}'$ and $\phim_j$ ($\phim_n = {\phim_n}'$ otherwise), we can Taylor expand $\hQ_{\{\phim_n\}, \{{\phim_n}'\}}$ in $\lambda_j$ as follows:
\begin{eqnarray}
    \hQ_{\{\phim_n\}, \{{\phim_n}'\}} &=& \left(\frac{1}{\mbP(\{\phim_n\})}\ket{\{\phim_n\}} - \frac{1}{\mbP(\{{\phim_n}'\})}\ket{\{{\phim_n}'\}}\right)\left(\frac{1}{\mbP(\{\phim_n\})}\bra{\{\phim_n\}} - \frac{1}{\mbP(\{{\phim_n}'\})}\bra{\{{\phim_n}'\}}\right) \nn \\
    &=& \lambda_j^2 \frac{\delta}{\delta \phim_j}\left(\frac{1}{\mbP(\{\phim_n\})}\ket{\{\phim_n\}}\right)\left(\frac{1}{\mbP(\{\phim_n\})}\bra{\{\phim_n\}}\right)\left(\frac{\delta}{\delta \phim_j}\right)^\dagger + \mathcal{O}\left(\lambda^3_j\right). 
\label{eq:Qtaylor}
\end{eqnarray}
Using Eq.~(\ref{eq:Hreg}) and (\ref{eq:Qtaylor}), we obtain
\begin{equation}
    H^{(m)} = \sigma^2\sumal{j}{}{\int{\prodal{n}{}{d \phim_n}\  \mbP(\{\phim_n\})^2 \frac{\delta}{\delta \phim_j}\left(\frac{1}{\mbP(\{\phim_n\})}\ket{\{\phim_n\}}\right)\left(\frac{1}{\mbP(\{\phim_n\})}\bra{\{\phim_n\}}\right)\left(\frac{\delta}{\delta \phim_j}\right)^\dagger + \mathcal{O}\left(\sigma^3\right)}}
\end{equation}
Using the fact that $\sigma^2 \sim \dx$, up to an overall dimensionful factor, the continuum Hamiltonian is
\begin{equation}
    H^{(m)} = \frac{\gamma}{2}\intal{}{}{dx\ \int{\mathcal{D}\phim\  \mbP(\phim(x))^2\ \frac{\delta}{\delta \phim(x)}\left(\frac{1}{\mbP(\phim(x))}\ket{\phim(x)}\right)\left(\frac{1}{\mbP(\phim(x))}\bra{\phim(x)}\right)\left(\frac{\delta}{\delta \phim(x)}\right)^\dagger}}.
\label{eq:HRKcont}
\end{equation}
Next, from Eqs.~(\ref{eq:params}) and (\ref{eq:probhigher}) we compute 
\begin{equation}
    \mbP(\phim(x)) \frac{\delta}{\delta \phim(x)}\left(\frac{1}{\mbP(\phim(x))} \ket{\phim(x)}\right) =  \left(\frac{\delta}{\delta\phim(x)} + \frac{\beta}{2} \frac{\delta \mE(\phim_k)}{\delta \phim(x)}\right)\ket{\phim(x)},
\label{eq:simp}
\end{equation}
and
\begin{eqnarray}
    &\frac{\delta \mE(\phim_k)}{\delta \phim(x)} = - \frac{\delta}{\delta\phim(x)}\ \int_0^L{dy\ \left(\partial^{m+1}_y \phim(y)\right)^2} = - 2\int_0^L{dy\ \partial^{m+1}_y \phim(y) \partial^{m+1}_y \delta(y - x)} \nn \\
    &= -2\times (-1)^{m+1} \partial^{2(m+1)}_x \phim(x),
\label{eq:Ederivative}
\end{eqnarray} 
where we have integrated by parts. 
Finally, using Eq.~(\ref{eq:simp}) we obtain the functional form of the Hamiltonian $H^{(m)}$
\begin{eqnarray}
    H^{(m)} &=& \frac{\gamma}{2}\int{dx\ \left(-\frac{\delta}{\delta \phim(x)} + (-1)^{m+1}\kappa\ \partial_x^{2(m+1)} \phim(x) \right)\left(\frac{\delta}{\delta \phim(x)} + (-1)^{m+1}\kappa\ \partial_x^{2(m+1)} \phim(x) \right)} \nn \\
    &=& \gamma\left(\int{dx\ \left(\frac{{\Pi^{(m)}}^2}{2} + \frac{\kappa^2}{2}\ (\partial_x^{2(m+1)} \phim(x))^2 \right)} + (-1)^{m+1}\frac{\kappa}{2} \int{dx\ \partial_x^{2(m+1)} \delta(x)}\right) \, ,
\label{eq:hsmffinal}
\end{eqnarray}
matching the expressions in Eqs.~(\ref{eq:hamilhm}), (\ref{eq:createannihildefn}) and (\ref{eq:continuumhamil}) in the main text, up to an overall constant energy shift. 
\subsection{Fokker-Planck approach}
\label{app:continuumhamilFP}
An alternate, more standard approach for deriving the continuum Hamiltonian proceeds by invoking an analogy with Langevin dynamics~\cite{henley1997relaxation}.
We begin with the observation that that the ground state wave functional of the continuum RK Hamiltonian---when interpreted as the equilibrium probability distribution of the Markov process in the continuum---given by the absolute square of the amplitudes in Eqs.~(\ref{eq:wf}) and (\ref{eq:notfreeparticle}), resembles a Boltzmann distribution
\begin{equation}
    W_0(\phim(x)) = \frac{1}{\mathcal{Z}}\exp(-\kappa \mE(\phim(x))) \, ,
\label{eq:boltzmanndist}
\end{equation}
where $\kappa$ is the ``inverse-temperature" and $\mE(\phim(x))$ is the energy of the system given by
\begin{equation}
    \mE(\phim(x)) = \int_0^L dx\ \left(\partial^{m+1}_x \phim(x)\right)^2.
\end{equation}
Given that the underlying Markov process relaxes to the Gibbs distribution, we expect the long-wavelength behavior at late-times to be captured by Langevin dynamics of the generalized height fields $\phim(x, t)$~\cite{chaikin1995principles, henley1997relaxation} 
\begin{equation}
    \frac{d \phim(x, t)}{d t} = -\frac{\gamma \kappa}{2} \frac{\delta \mE(\phim(x, t))}{\delta \phim(x)} + \zeta(x, t) \, .
\label{eq:langevin}
\end{equation}
Here, $\gamma$ is a constant that sets the overall rate of relaxation and $\zeta(x, t)$ is $\delta$-correlated ``white-noise" i.e., it satisfies a Gaussian distribution with
\begin{equation}
    \langle \zeta(x, t) \rangle = 0,\;\;\; \langle \zeta(x, t) \zeta(x', t') \rangle = \gamma\ \delta(x - x')\delta(t - t') \, .
\label{eq:whitenoiseproperties}
\end{equation}
To see that the distribution Eq.~(\ref{eq:boltzmanndist}) is indeed the equilibrium distribution of Eq.~(\ref{eq:langevin}), we derive the time-evolution equation for the joint probability densities, also known as the Fokker-Planck equation.
We follow methods elucidated in Refs.~\cite{henley1997relaxation, henley2004, chaikin1995principles, Risken1996}.
Note that although we consider OBC in the main text, we will use PBC in the following for convenience.
We first define Fourier transforms of the height fields $\{\phim_k\}$ as follows
\begin{equation}
    \phim(x) = \frac{1}{\sqrt{L}}\sumal{k \neq 0}{}{e^{ikx}\phim_k},\;\;\phim_k = \frac{1}{\sqrt{L}}\int_0^L{dx\ e^{-ikx}\phim(x)}\, ,\;\;\mE(\phim_k) \equiv \sumal{k \neq 0}{}{k^{2(m+1)} \phim_k \phim_{-k}}\, , 
\label{eq:Ephi}
\end{equation}
where $k$ is a discrete momentum variable, and we have suppressed the dependence on $t$ and without loss of generality set $\phim_{k = 0} = 0$.
Taking the the Fourier transform of Eq.~(\ref{eq:langevin}), we then obtain
\begin{equation}
    \frac{1}{\sqrt{L}}\sumal{k \neq 0}{}{e^{ikx}\frac{d\phim_k}{dt}} =  -\frac{\gamma\kappa}{2} \sumal{k \neq 0}{}{\frac{d \mE(\phim_k) }{d \phim_{-k}}\frac{\delta\phim_{-k}}{\delta \phim(x)}} + \frac{1}{\sqrt{L}}\sumal{k \neq 0}{}{e^{ikx}\zeta_k(t)}\nn \implies \frac{d\phim_k}{dt} =  -\frac{\gamma \kappa}{2}\frac{d \mE(\phim_k)}{d\phim_{-k}} + \zeta_k(t). 
\label{eq:langevinFT}
\end{equation}
Further, using Eqs.~(\ref{eq:whitenoiseproperties}) and~(\ref{eq:langevinFT}), we obtain
\begin{equation}
    \langle \zeta_k(t)\rangle = 0,\;\;\;\langle \zeta_k(t) \zeta_{k'}(t')\rangle = \gamma\ \delta_{k'
    , -k} \delta(t - t') \, .
\label{eq:FTvals}
\end{equation}
Following standard methods in Langevin theory~\cite{chaikin1995principles, Risken1996}, we write the expression for the probability distribution at time $t + \dt$ as
\begin{equation}
    W(\phim_k, t + \dt) = \int{\mathcal{D}{\phim}' P(\phim_k, t + \dt | {\phim_k}', t)\ W({\phim_k}', t)},\;\;\;\mathcal{D}{\phim}'\equiv \prodal{k\neq 0}{}{d{\phim_k}'}
\label{eq:propagator}
\end{equation}
where $W(\phim_k , t)$ represents the joint probability distribution of the height variables $\{\phim_k\}$ at time $t$, and $P(\phim_k, t + \dt | {\phim_k}', t)$ denotes the probability of transition of the height variables from $\{{\phim_k}'\}$ to $\{\phim_k\}$ in time $\dt$. 
Defining
\begin{equation}
    \Delta\phim_k \equiv \phim_k - {\phim_k}', 
\label{eq:dphidefn}
\end{equation}
we rewrite Eq.~(\ref{eq:propagator}) as
\begin{equation}
    W(\phim_k, t + \dt) = \int{\mathcal{D}\left({\Delta\phim}\right) P(\phim_k, t + \dt | \phim_k - \Delta\phim_k, t)\ W(\phim_k - \Delta\phim_k, t)},\;\;\;\mathcal{D}\left({\Delta\phim}\right) \equiv \prodal{k\neq0}{}{d(\Delta\phim_k)}.
\label{eq:propagatorrewrite}
\end{equation}
We then Taylor-expand the integrand in Eq.~(\ref{eq:propagatorrewrite}) as
\begin{eqnarray}
    &P(\phim_k - \Delta\phim_k + \Delta\phim_k, t + \dt | \phim_k - \Delta\phim_k, t)\ W(\phim_k - \Delta\phim_k, t) \nn \\
    &=  \left(1  - \sumal{k \neq 0}{}{\Delta\phim_k \frac{d}{d\phim_k}} + \frac{1}{2}\sumal{k, k' \neq 0}{}{\Delta\phim_k \Delta\phim_{k'}\frac{d^2}{d\phim_k d\phim_{k'}}} + \cdots\right)P(\phim_k + \Delta\phim_k, t + \dt | \phim_k, t)\ W(\phim_k, t).
\label{eq:integrandchange}
\end{eqnarray}
Defining the quantities
\begin{eqnarray}
    \langle \Delta\phim_k \rangle &\equiv& \int{\mathcal{D}(\Delta\phim) \Delta\phim_k P(\phim_k + \Delta\phim_k, t + \dt| \phim_k, t)} \nn \\
    \langle \Delta\phim_k \Delta\phim_{k'} \rangle &\equiv& \int{\mathcal{D}(\Delta\phim) \Delta\phim_k \Delta\phim_{k'} P(\phim_k + \Delta\phim_k, t + \dt| \phim_k, t)},
\label{eq:averagedefn}
\end{eqnarray}
we can write the Kramers-Moyal expansion~\cite{Risken1996} of Eq.~(\ref{eq:propagatorrewrite}) as  
\begin{equation}
    W(\phim_k, t + \dt) =  \left(1  - \sumal{k \neq 0}{}{ \frac{d}{d\phim_k}} \langle \Delta\phim_k \rangle + \frac{1}{2}\sumal{k, k' \neq 0}{}{\frac{d^2}{d\phim_k d\phim_{k'}} \langle \Delta\phim_k \Delta\phim_{k'} \rangle} + \cdots\right)\ W(\phim_k, t),
\label{eq:KMexpansion}
\end{equation}
where the derivatives also act on $W$.
Using Eqs.~(\ref{eq:langevinFT}) and (\ref{eq:averagedefn}), we find that
\begin{align}
    \langle \Delta\phim_k \rangle &= -\frac{\gamma \kappa}{2} \frac{d \mE(\phim_k)}{d \phim_{-k}}\dt + \int_t^{t + \dt}{dt'\ \langle \zeta_k(t') \rangle} = - \gamma \kappa k^{2(m + 1)}\phim_k\dt \, , \nn \\
    \langle \Delta\phim_k \Delta\phim_{k'} \rangle &= \gamma \dt\ \delta_{k', -k} + \mathcal{O}((\dt)^2) \, ,
\label{eq:langevininfinitesimal}
\end{align}
where we have used Eq.~(\ref{eq:FTvals}). 
Using Eqs.~(\ref{eq:KMexpansion}) and \eqref{eq:langevininfinitesimal},  we obtain
\begin{equation}
    \frac{d W(\phim_k, t)}{dt} = \frac{\gamma}{2} \sumal{k \neq 0}{}{\frac{d}{d\phim_k}\left(\frac{d}{d \phim_{-k}} + 2 \kappa k^{2(m+1)} \phim_k\right)}\ W(\phim_k, t).
\label{eq:mastereq}
\end{equation}
Note that Eq.~(\ref{eq:mastereq}) is a master equation of the form of Eq.~(\ref{eq:master_eq_matrix}).
We can then directly verify that the equilibrium probability distribution is given by
\begin{equation}
    W_0(\phim_k) = \frac{1}{\mathcal{Z}}\exp\left(- \kappa \sumal{k \neq 0}{}{k^{2(m + 1)}\phim_k \phim_{-k}}\right),
\label{eq:boltzmannft}
\end{equation}
which is the Fourier transform of the Boltzmann distribution Eq.~(\ref{eq:boltzmanndist}).
To obtain an symmetric (SMF decomposable) quantum Hamiltonian corresponding to the master equation Eq.~(\ref{eq:mastereq}) with the wavefunction of Eq.~(\ref{eq:wf}) as the ground state, we perform a similarity transformation on the transition matrix in Eq.~(\ref{eq:mastereq}); equivalently, we write Eq.~(\ref{eq:mastereq}) in the form~\cite{henley1997relaxation, henley2004, castelnovo2005}
\begin{equation}
    \frac{d\Phi^{(m)}(\phim_k, t)}{d t} = - H^{(m)} \Phi^{(m)}(\phim_k, t),\;\;\; \Phi^{(m)}(\phim_k, t) = \frac{W(\phim_k, t)}{\sqrt{W_0(\phim_k)}},
\label{eq:FPlangevin}
\end{equation}
which, using Eqs.~(\ref{eq:mastereq}) and the form of Eq.~(\ref{eq:boltzmannft}), can be shown to be~\cite{henley1997relaxation}
\begin{equation}
    \frac{d\Phi^{(m)}(\phim_k, t)}{d t} = -\frac{\gamma}{2} \sumal{k \neq 0}{}{\left(-\frac{d}{d\phim_k} + \kappa k^{2(m+1)} \phim_{-k}\right)\left(\frac{d}{d\phim_{-k}} + \kappa k^{2(m+1)} \phim_{k}\right)}\Phi^{(m)}(\phim_k, t). 
\label{eq:SMFHamilcontinuum}
\end{equation}
Thus, we find that the continuum Hamiltonian $H^{(m)}$ is given by
\begin{equation}
    H^{(m)} = \gamma \sumal{k \neq 0}{}{\mQ^\dagger_m(k) \mQ_m(k)}, 
\label{eq:hsmfkspace}
\end{equation}
where the creation and annihilation operators are defined as
\begin{equation}
    \mQ^\dagger_m(k) = \frac{1}{\sqrt{2}}\left(-\frac{d}{d\phim_k} + \kappa k^{2(m + 1)} \phim_{-k}\right),\;\;\mQ_m(k) = \frac{1}{\sqrt{2}}\left(\frac{d}{d\phim_{-k}} + \kappa k^{2(m + 1)} \phim_{k}\right)
\label{eq:createannihil}    
\end{equation}
and satisfy the algebra
\begin{equation}
    \left[\mQ_m(k), \mQ_m(k')\right] = 0,\;\;\;\left[\mQ^\dagger_m(k), \mQ^\dagger_m(k')\right] = 0,\;\;\;\left[\mQ_m(k), \mQ^\dagger_m(k')\right] = \kappa\ k^{2(m+1)} \delta_{k, k'}.
\label{eq:algebra}
\end{equation}
As a direct consequence of the construction, we can verify that the ground state wavefunction $\Phi^{(m)}_0(\phim_k)$ of $H^{(m)}$ is annihilated by all $\mQ(k)$ and is given by
\begin{equation}
    \Phi^{(m)}_0(\phim_k) = \frac{1}{\sqrt{\mZ}}\exp\left(-\frac{\kappa}{2}\sumal{k \neq 0}{}{k^{2(m+1)} \phim_k \phim_{-k}}\right),\;\;\; H^{(m)}\Phi^{(m)}_0(\phim) = 0.
\label{eq:groundstatewf}
\end{equation} 
Similarly, the excited state wavefunction $\Phi^{(m)}_k(\phim_k)$ of momentum $k$ can be constructed by acting the creation operator $\mQ^\dagger_m(k)$ on the ground state 
\begin{equation}
    \Phi^{(m)}_k(\phim_k) = \mQ^\dagger_m(k) \Phi^{(m)}_0(\phim_k),\;\;\; H^{(m)}\Phi^{(m)}_k(\phim_k) = \gamma \kappa\ k^{2(m+1)}\Phi^{(m)}_k(\phim_k). 
\label{eq:excitedstatewf}
\end{equation}
We can also express the Hamiltonian of Eq.~(\ref{eq:algebra}) in terms of real-space creation and annihilation operators $\mQ^\dagger_m(x)$ and $\mQ_m(x)$ respectively as
\begin{eqnarray}
    &H^{(m)} = \gamma \int_0^L{dx\ \mQ^\dagger_m (x) \mQ_m(x)} \nn \\
    &\mQ^\dagger_m(x) = \frac{1}{\sqrt{L}}\sumal{k \neq 0}{}{dx\ e^{-i k x} \mQ^\dagger_m (k)} = \frac{1}{\sqrt{2}}\left(- \frac{\delta}{\delta \phim(x)} + (-1)^{m+1} \kappa\ \partial^{2(m+1)}_x \phim(x)\right),\nn \\
    &\mQ_m(x) = \frac{1}{\sqrt{L}}\sumal{k \neq 0}{}{dx\ e^{i k x} \mQ_m (k)} = \frac{1}{\sqrt{2}}\left( \frac{\delta}{\delta \phim(x)} + (-1)^{m+1} \kappa\ \partial^{2(m+1)}_x \phim(x)\right),
\label{eq:realspaceops}
\end{eqnarray}
which obey the commutation relations
\begin{equation}
    \left[\mQ_m(x), \mQ_m(y)\right] = 0,\;\;\;\left[\mQ^\dagger_m(x), \mQ^\dagger_m(y)\right] = 0,\;\;\;\left[\mQ_m(x), \mQ^\dagger_m(y)\right] = (-1)^{m+1} \kappa\ \partial_x^{2(m+1)} \delta(x - y).
\label{eq:realspacealgebra}
\end{equation}
Consequently, the ground state wavefunction is annihilated by $\mQ_m(x)$ and is given by
\begin{equation}
    \Phi^{(m)}_0(\phim(x)) = \frac{1}{\sqrt{\mZ}}\exp\left(-\frac{\kappa}{2}\int_0^L dx\ (\partial^{m+1}_x \phim(x))^2\right).
\end{equation}
Furthermore, following Eq.~(\ref{eq:excitedstatewf}), the excited wavefunction $\Phi^{(m)}_k(\phim(x))$ reads
\begin{equation}
    \Phi^{(m)}_k(\phim(x)) = \int_0^L{dx\ e^{i k x} \mQ^{\dagger} _m(x) \Phi^{(m)}_0(\phim(x))}.
\label{eq:excitedwfrealspace}
\end{equation}
\section{Continuum Hamiltonian in higher dimensions}\label{app:continuumhamilhigherd}
In this Appendix, we briefly sketch the derivation of the continuum Hamiltonian for $m^{th}$ multipole conserving systems in $d$-dimensions, following the Fokker-Planck approach illustrated in Appendix~\ref{app:continuumhamilFP}. 
As discussed in Sec.~\ref{sec:higherdims}, the ground state wavefunctional of the continuum RK-Hamiltonian, when interpreted as the equilibrium probability distribution of the Markov process in the continuum, is given by the absolute square of the amplitudes in Eq.~(\ref{eq:GSwfhigherd}):
\begin{equation}
    W_0(\Em{j}(\vx)) = \frac{1}{\mathcal{Z}}\exp(-\kappa \mE(\Em{j}(\vx))) \, ,
\label{eq:boltzmanndisthigherd}
\end{equation}
where $\kappa$ is the ``inverse-temperature" and $\mE(\Em{j}(\vx))$ is the ``energy" of the system, given by
\begin{equation}
    \mE(\Em{j}(\vx)) = \int d^d\vx\ \left(\partial_{j_0} \cdots \partial_{j_m} \Em{j}(\vx)\right)^2 ,
\end{equation}
where repeated indices are summed over. 
Similar to Eq.~(\ref{eq:langevin}) in the one dimensional case, we expect the long-wavelength behavior at late-times to be captured by Langevin dynamics of the generalized tensor fields $\Em{j}(\vx, t)$:
\begin{equation}
    \frac{d \Em{l}(\vx, t)}{d t} = -\frac{\gamma \kappa}{2} \frac{\delta \mE(\Em{j}(\vx, t))}{\delta \Em{l}(\vx)} + \zm{l}(\vx, t) \, ,
\label{eq:langevinhigherd}
\end{equation}
where $\zm{l}(\vx, t)$ is $\delta$-correlated ``white-noise" governing the dynamics of the component $\Em{l}(\vx, t)$ of the tensor field i.e., it satisfies a Gaussian distribution with
\begin{equation}
    \langle \zm{l}(\vx, t) \rangle = 0,\;\;\; \langle \zm{l}(\vx, t) \zm{p}(\vx', t') \rangle = \gamma \delta(\vx - \vx')\delta(t - t') \delta_{\{l_0 \cdots l_m\}, \{p_0 \cdots p_m\}} \, ,
\label{eq:whitenoisepropertieshigherd}
\end{equation}
where $\delta_{\{l_0 \cdots l_m\}, \{p_0 \cdots p_m\}}$ is $0$ unless the sets $\{l_0, \cdots, l_m\}$ and $\{p_0, \cdots, p_m\}$ are equal --- this ensures that the fluctuations of different components of the tensor field $\{\Em{j}\}$ are not correlated, as discussed in Sec.~\ref{sec:higherdims}.
Considering a system in $d$-dimensions with linear-size $L$ in each direction, we define Fourier transforms of the generalized tensor fields $\{\Em{j}_{\vk}\}$ as follows
\begin{eqnarray}
    &\Em{j}(\vx) = {L^{-\frac{d}{2}}}\sumal{\vk \neq 0}{}{e^{i\vk\cdot \vx} \Em{j}_{\vk}},\;\;\;
    \Em{j}_{\vk} = L^{-\frac{d}{2}}\int{d^d \vx\ e^{-i\vk\cdot \vx}\phim(x)}\, ,\nn \\
    &\mE(\Em{j}_{\vk}) \equiv \sumal{\vk \neq 0}{}{(k_{i_0}\cdots k_{i_m} \Em{i}_{\vk}) (k_{j_0}\cdots k_{j_m} \Em{j}_{-\vk})}\, \label{eq:Ephi}
\end{eqnarray}
where repeated indices are summed over, $\vk$ is a discrete momentum variable, and we set $\Em{j}_{\vk = 0} = 0$ without loss of generality.
Eqs.~(\ref{eq:langevinhigherd}) and (\ref{eq:whitenoisepropertieshigherd}) can then be written as
\begin{align}
   \frac{d\Em{l}_{\vk}}{dt} =  -\frac{\gamma \kappa}{2} & \frac{d \mE(\Em{j}_{\vk})}{\Em{l}_{-\vk}} +  \zm{l}_{\vk}(t) \, , \nn \\
   \langle \zm{l}_{\vk}(t)\rangle = 0,\;\;\;  \langle \zm{l}_{\vk}(t) \zm{p}_{\vk'}(t')\rangle = & \, \gamma \delta_{\vk'
    , -\vk} \delta(t - t') \delta_{\{l_0, \cdots, l_m\}, \{p_0, \cdots, p_m\}}.
\label{eq:langevinFThigherd}
\end{align}
We further obtain a Fokker-Planck master equation for the probability distribution of generalized tensor fields $W(\Em{j}_{\vk}, t)$ starting from Eq.~(\ref{eq:langevinFThigherd}) (following steps similar to Eqs.~(\ref{eq:propagator})-(\ref{eq:mastereq}) in the one-dimensional case) and subsequently perform a similarity transformation to obtain the expression for the SMF decomposable Hamiltoian $H^{(m)}$ (following steps similar to Eqs.~(\ref{eq:FPlangevin})-(\ref{eq:SMFHamilcontinuum})).
Ultimately, $H^{(m)}$ can be shown to be of the form
\begin{eqnarray}
    &H^{(m)} = \gamma \sumal{\vk \neq 0}{}{ (\mQ^\dagger_m(\vk))^{l_0 \cdots l_m} (\mQ_m(\vk))_{l_0 \cdots l_m}},\nn \\
    &(\mQ^\dagger_m(\vk))^{l_0 \cdots l_m} = \frac{1}{\sqrt{2}}\left(-\frac{d}{d \Em{l}_{\vk}} + \kappa\ k^{l_0} \cdots k^{l_m} (k_{j_0} \cdots k_{j_m} \Em{j}_{-\vk})\right),\nn \\
    &(\mQ_m(\vk))_{l_0 \cdots l_m} = \frac{1}{\sqrt{2}}\left(\frac{d}{d\Em{l}_{-\vk}} + \kappa\ k_{l_0} \cdots k_{l_m} (k_{j_0} \cdots k_{j_m} \Em{j}_{\vk})\right),
\label{eq:createannihilhigherd}    
\end{eqnarray}
where repeated indices are summed over, and we do not distinguish between upper and lower indices (i.e. $k^l = k_l$). 
The creation and annilation operators satisfy the algebra
\begin{eqnarray}
    &\left[(\mQ_m(\vk))_{l_0 \cdots l_m}, (\mQ_m(\vk'))_{p_0 \cdots p_m}\right] = 0,\;\;\;\left[(\mQ^\dagger_m(\vk))^{l_0 \cdots l_m}, (\mQ^\dagger_m(\vk'))^{p_0 \cdots p_m}\right] = 0,\nn \\
    &\left[(\mQ_m(\vk))_{l_0 \cdots l_m}, (\mQ^\dagger_m(\vk'))^{p_0 \cdots p_m}\right] = \kappa\ (k_{l_0} \cdots k_{l_m}) (k^{p_0} \cdots k^{p_m})\ \delta_{\vk, \vk'}\;\;\;\forall l_j, p_j,\;\;0 \leq j \leq m, 
\label{eq:algebrahigherd}
\end{eqnarray}
where we do not sum over repeated indices in the second line. 
Consequently, the ground state wavefunction $\Phi^{(m)}_0(\Em{j}_{\vk})$ of $H^{(m)}$ is annihilated by all $(\mQ_m(\vk))_{l_0 \cdots l_m}$ and is given by
\begin{equation}
    \Phi^{(m)}_0(\Em{j}_{\vk}) = \frac{1}{\sqrt{\mZ}}\exp\left(-\frac{\kappa}{2}\sumal{k \neq 0}{}{(k_{j_0}\cdots k_{j_m} \Em{j}_{\vk})(k_{l_0}\cdots k_{l_m} \Em{l}_{-\vk})}\right),\;\;\; H^{(m)}\Phi^{(m)}_0(\Em{j}_{\vk}) =0. 
\label{eq:groundstatehigherd}
\end{equation}
Similarly, the excited state wavefunction $\Phi^{(m)}_{\vk}(\Em{j}_{\vk})$ of momentum $\vk$ can be constructed by acting a superposition of the creation operators $\{(\mQ^\dagger_m(\vk))^{l_0 \cdots l_m}\}$ on the ground state 
\begin{equation}
    \Phi^{(m)}_{\vk}(\Em{j}_{\vk}) = (k_{l_0} \cdots k_{l_m}(\mQ^\dagger_m(\vk))^{l_0 \cdots l_m}) \Phi^{(m)}_0(\Em{j}_{\vk}),\;\;\; H^{(m)}\Phi^{(m)}_{\vk}(\Em{j}_{\vk}) = \gamma \kappa\ (k_{l} k^{l})^{m+1}\Phi^{(m)}_{\vk}(\Em{j}_{\vk}), 
\label{eq:excitedstatewfhigherd}
\end{equation}
where repeated indices are summed over.
We can also express the Hamiltonian Eq.~(\ref{eq:algebrahigherd}) in terms of real-space creation and annihilation operators $\{(\mQ^\dagger_m(\vx))^{l_0 \cdots l_m}\}$ and $\{(\mQ_m(\vx))_{l_0 \cdots l_m}\}$ respectively as
\begin{eqnarray}
    &H^{(m)} = \gamma \int{d^d\vx\ (\mQ^\dagger_m(\vx))^{l_0 \cdots l_m} (\mQ_m(\vx))_{l_0 \cdots l_m}} \nn \\
    &(\mQ^\dagger_m(\vx))^{l_0 \cdots l_m} = L^{-\frac{d}{2}}\sumal{\vk \neq 0}{}{d^d\vx\ e^{-i \vk\cdot\vx} (\mQ^\dagger_m(\vk))^{l_0 \cdots l_m}} = \frac{1}{\sqrt{2}}\left(- \frac{\delta}{\delta E^{l_0 \cdots l_m}(\vx)} + (-1)^{m+1} \kappa\ \partial^{l_0} \cdots \partial^{l_m} (\partial_{j_0} \cdots \partial_{j_m} \Em{j}(\vx))\right),\nn \\
    &(\mQ_m(x))_{l_0 \cdots l_m} = L^{-\frac{d}{2}}\sumal{\vk \neq 0}{}{d^d\vx\ e^{i \vk \cdot \vx} (\mQ_m (\vk))_{l_0 \cdots l_m}} = \frac{1}{\sqrt{2}}\left( \frac{\delta}{\delta E^{l_0 \cdots l_m}(\vx)} + (-1)^{m+1} \kappa\ \partial_{l_0} \cdots \partial_{l_m} (\partial_{j_0} \cdots \partial_{j_m} \Em{j}(\vx))\right),\nn \\
\label{eq:realspaceopshigherd}
\end{eqnarray}
where we do not distinguish between upper and lower indices in the derivative operator (i.e. $\partial_l = \partial^l$). 
Following Eq.~(\ref{eq:algebrahigherd}), the creation and annihilation operators obey the commutation relations
\begin{eqnarray}
    &\left[(\mQ_m(\vx))_{l_0 \cdots l_m}, (\mQ_m(\vy))_{p_0 \cdots p_m}\right] = 0,\;\;\;\left[(\mQ^\dagger_m(\vx))^{l_0 \cdots l_m}, (\mQ^\dagger_m(\vy))^{p_0 \cdots p_m}\right] = 0,\nn \\
    &\left[(\mQ_m(\vx))_{l_0 \cdots l_m}, (\mQ^\dagger_m(\vy))^{p_0 \cdots p_m}\right] = (-1)^{m+1} \kappa\  \partial_{l_0} \cdots \partial_{l_m} \partial^{p_0} \cdots \partial^{p_m}\delta(\vx - \vy).
\label{eq:realspacealgebrahigherd}
\end{eqnarray}
Consequently, the ground state wavefunction is annihilated by $(\mQ_m(\vx))_{l_0 \cdots l_m}$ and is given by
\begin{equation}
    \Phi^{(m)}_0(\Em{j}(\vx)) = \frac{1}{\sqrt{\mZ}}\exp\left(-\frac{\kappa}{2}\int d^d\vx\ (\partial_{j_0} \cdots \partial_{j_m} \Em{j}(\vx))^2\right).
\end{equation}
Furthermore, following Eq.~(\ref{eq:excitedstatewfhigherd}), the excited wavefunction $\Phi^{(m)}_{\vk}(\Em{j}(\vx))$ reads
\begin{equation}
    \Phi^{(m)}_{\vk}(\Em{j}(\vx)) = \int{d^d\vx\ e^{i \vk\cdot \vx} (k_{l_0} \cdots k_{l_m})(\mQ^{\dagger} _m(\vx))^{l_0 \cdots l_m} \Phi^{(m)}_0(\Em{j}(\vx))}.
\label{eq:excitedwfrealspacehigherd}
\end{equation}
\end{document}